\documentclass[%
 reprint,
 amsmath,amssymb,
 aps,
prb,
]{revtex4-2}

\usepackage{xcolor}
\usepackage{graphicx}
\usepackage{dcolumn}
\usepackage{bm}
\usepackage[colorlinks = true,
            linkcolor = blue,
            urlcolor  = blue,
            citecolor = blue,
            anchorcolor = blue]{hyperref}

\begin{document}

\preprint{APS/123-QED}

\title{Simplicity of mean-field theories in neural quantum states}

\author{Fabian Ballar Trigueros}
\author{Tiago Mendes-Santos}
\author{Markus Heyl}%
\affiliation{%
 Theoretical Physics III, Center for Electronic Correlations and Magnetism,
Institute of Physics, University of Augsburg, 86135 Augsburg, Germany
}%

\date{\today}

\begin{abstract}
The utilization of artificial neural networks for representing quantum many-body wave functions has garnered significant attention, with enormous recent progress for both ground states and non-equilibrium dynamics. However, quantifying state complexity within this neural quantum states framework remains elusive. In this study, we address this key open question from the complementary point of view: Which states are simple to represent with neural quantum states? Concretely, we show on a general level that ground states of mean-field theories with permutation symmetry only require a limited number of independent neural network parameters. We analytically establish that, in the thermodynamic limit, convergence to the ground state of the fully-connected transverse-field Ising model (TFIM), the mean-field Ising model, can be achieved with just one single parameter. Expanding our analysis, we explore the behavior of the 1-parameter ansatz under breaking of the permutation symmetry. For that purpose, we consider the TFIM with tunable long-range interactions, characterized by an interaction exponent $\alpha$. We show analytically that the 1-parameter ansatz for the neural quantum state still accurately captures the ground state for a whole range of values for $0\le \alpha \le 1$, implying a mean-field description of the model in this regime.

\end{abstract}

\maketitle


\section{\label{sec:Intro}Introduction}

Representing the wave function of complex quantum matter is exceedingly difficult. Addressing this challenge has prompted the proposal of various techniques and approximations \cite{PhysRev.136.B864, PhysRevLett.93.076401, PhysRevLett.93.040502, PhysRevLett.69.2863,RevModPhys.87.1067,RevModPhys.73.33}. However, each method encounters distinct difficulties. For instance, exact diagonalization becomes computationally intractable for large systems due to the exponentially growing basis. Quantum Monte Carlo methods can encounter the sign problem \cite{PhysRevB.41.9301} for many interesting systems. Tensor networks experience an exponentially increasing bond dimension for systems with volume-law scaling of entanglement entropy \cite{10.5555/2011832.2011833,PhysRevB.73.094423}. Recently, neural quantum states (NQSs) have been introduced as an alternative method to leverage the expressive power of neural networks to represent the quantum wave function \cite{Carleo_2017}. 

This approach has delivered remarkable results in discovering ground states and describing dynamics of quantum many-body systems in previously inaccessible regimes via other methods \cite{PhysRevLett.125.100503, chen2023efficient, Mendes_Santos_2023, doi:10.1073/pnas.2122059119, PhysRevB.100.125124}. One key feature that distinguishes NQS from tensor networks is the ability to represent volume-law entangled states \cite{PhysRevX.7.021021}. 

Despite demonstrations of the expressive power of NQS \cite{Carleo_2018,PhysRevB.106.205136}, there is a critical need to understand a metric of complexity for it, similar to how bond dimension and circuit depth are utilized for characterizing tensor networks and quantum circuits, respectively. The outstanding question is: What criteria can be employed to quantify the complexity of a neural quantum state? In a first step we approach this question from the complementary side: Which states can be easily described by a neural quantum network? Much like how we know that area-law states can be efficiently described using a tensor network, recognizing the states that NQS can easily describe might be crucial in advancing our understanding of NQS complexity.

A natural choice is to consider that the complexity is related to the number of independent variational parameters, $K$, a trained NQS requires to describe a specific quantum state accurately. In this work, we characterize states that have minimal complexity in this sense under the neural quantum states formalism. In particular, we show that neural quantum states can describe mean-field theories with permutation symmetry with a very small number of parameters. We show that on a general level, this leads to a reduction in required network parameters from $K \times L$ to just $K$, where $K$ and $L$ correspond to the number of hidden neurons and the number of physical spins respectively, see Fig.~\ref{fig:RBMilust} for an illustration.

We demonstrate this using the fully-connected transverse-field Ising model. Most importantly we find that for this model convergence in the thermodynamic limit is achieved even with a single parameter, i.e., $K=1$. This makes the mean-field solution of the TFIM as simple as possible for NQS. As a next step, we study the behavior of the network beyond mean-field theory by breaking the permutation symmetry. For that purpose, we consider the long-range interacting TFIM with power-law decaying interactions. This allows us to tune the deviation from mean-field through the interaction exponent $\alpha$ in a controlled way. We show analytically that a neural network with a single parameter is sufficient to obtain the ground state for $\alpha \le 1$ in the thermodynamic limit. As a consequence, we find that the ground state of the long-range interacting TFIM is still described by a MFT for this range of values of $\alpha$.

\section{\label{sec:Permut} Neural quantum states and permutation invariance}

\begin{figure*}
    \centering
    \includegraphics[width = 0.6\linewidth]{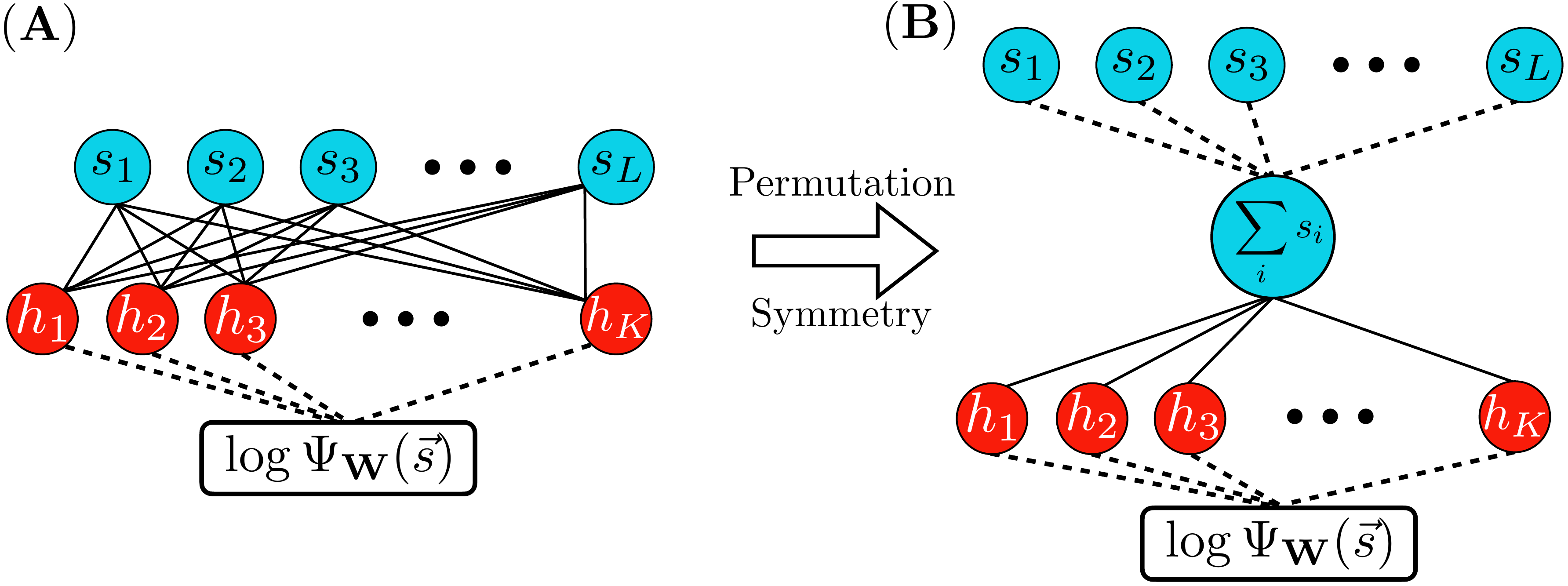}
    \includegraphics[width=0.38\linewidth]{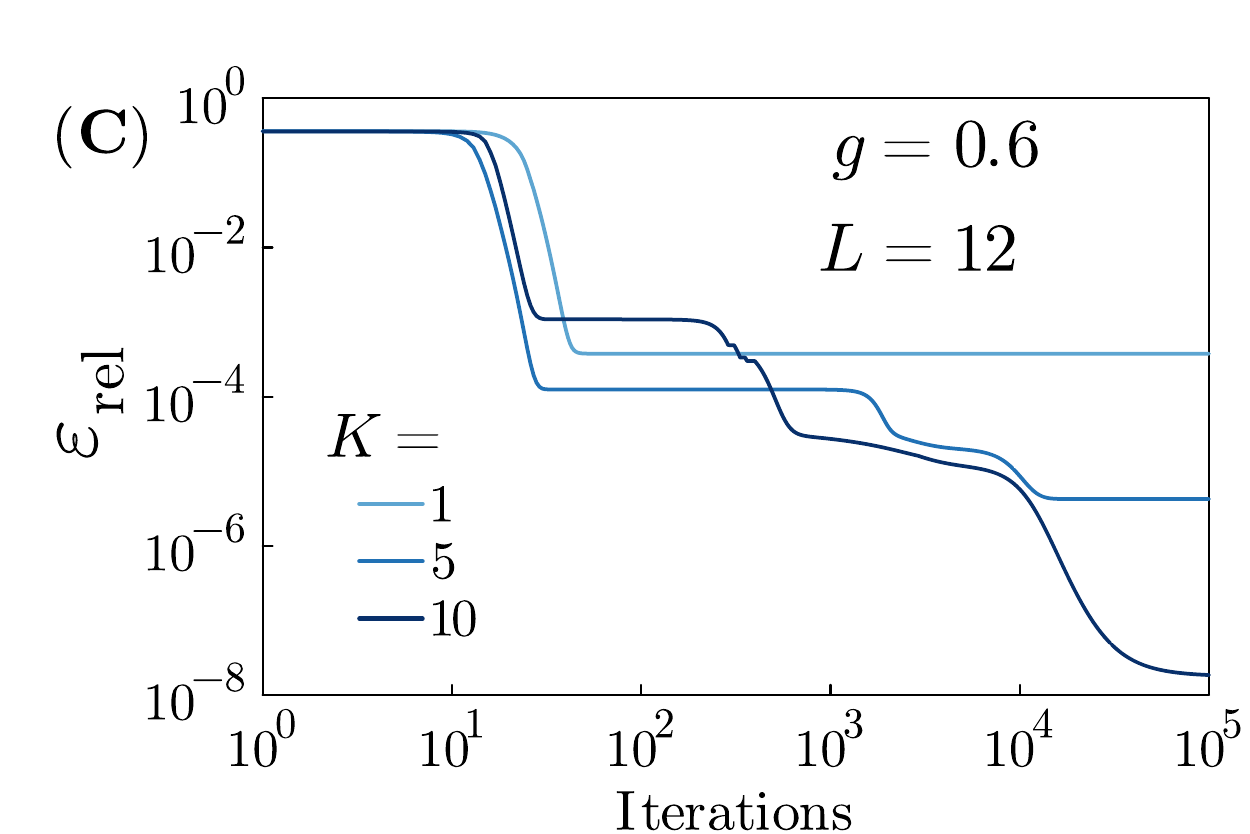}
    \caption{Permutation-invariant artificial neural networks and training.  $(\bf{A})$ An unconstrained structure of a general feed-forward network with one hidden layer. Dashed lines carry no variational parameters whereas full lines denote the fact that we compute the product between the input spin and a variational parameter. $(\bf{B})$ Enforcing permutation symmetry leads effectively to a modified neural network architecture, where the input spin configuration is first transformed into a collective total spin before further processing occurs. $(\bf{C})$ Training convergence and scaling of the ground state search process for the fully-connected transverse-field Ising model for different numbers of hidden spins $K$ at a fixed system size $L=12$. Here, $\epsilon_{\textrm{rel}}$ is the relative energy error of the neural quantum state with the exact diagonalization result as a reference. }
    \label{fig:RBMilust}
\end{figure*}

Neural quantum states offer a framework that leverages the expressive capabilities and generalization power of neural networks for representing quantum wave functions. This approach relies around the notion of expressing a wave function within a complete basis set denoted as $|s\rangle$, characterized by the expansion:

\begin{equation}
|\Psi\rangle = \sum_{s}^{} \Psi_\textbf{W}(s) |s\rangle.
\end{equation}

\noindent
Here, $\Psi_\textbf{W}(s)$ represents the neural network ansatz which depends on a set of variational parameters $\textbf{W}$ that we learn according to a prescription of choice. In this work we will consider a 1D lattice of size $L$ with periodic boundary conditions and spin-$1/2$ degrees of freedom. We take as a basis $|s\rangle$, the spin configuration in the computational basis.

We are particularly interested in the structure of $\textbf{W}$ for a permutation invariant Hamiltonian as they appear naturally for mean-field theories. For this, we will approximate the wave function by an artificial neural network. So we consider a general feed-forward neural network with $L$-input units, $K$-hidden neurons, no bias, and lastly a single output neuron that returns the value of $\log ( \Psi_\textbf{W}(s) )$. A schematic depiction of the initial architecture can be found in Fig.\,\ref{fig:RBMilust}\textbf{A}.

One important technique to minimize the number of parameters needed to describe the wave function in NQS is to take advantage of the symmetries at the level of the weight matrix \cite{Carleo_2017}, reducing the number of independent parameters. As a consequence, given that the network doesn't need to find the symmetry on its own, the computational cost of NQS can be greatly reduced without compromising the accuracy of the model.

The variational ansatz for our architecture can be expressed as:

\begin{equation}
    \Psi_{\textbf{W}}(s) = \langle s | \Psi \rangle = \exp \left[ \sum_{i}^{} f\left( y_{i}(s) \right) \right],
\end{equation}

\noindent
where $y_{i}(s) \equiv \sum_{j}^{} W_{i j} s_j$, $f$ denotes the activation function and $W_{i j}$ represents the weight matrix elements. However, to ensure that the output remains invariant under any permutation of the input configuration, we impose a constraint between the elements of $W_{i,j}$. For this, we consider the set of all possible permutations of the form $\pi : {s_j} \rightarrow {s_{\pi(j)}}$, i.e., $\pi(s) \equiv (s_{\pi(1)},\dots,s_{\pi(L)})$ and we require that

\begin{equation}
y_i(s) - y_{i}(\pi(s)) = \sum_{j} W_{i j} ( s_j - \pi(s)_j) \overset{!}{=}  0.
\end{equation}

\noindent
Given that due to the nature of permutations $\pi$ is a bijective mapping and hence there is a unique $\pi^{-1}$, we then rewrite our constraint in a more convenient way as,

\begin{equation}
     \sum_{j}( W_{i j} - W_{i \pi^{-1}(j)})s_j \overset{!}{=}  0.
\end{equation}

\noindent
Since this is supposed to hold for any permutation and to be independent of the values of $s_j$, it follows that the quantity in parenthesis has to vanish, e.g., $ W_{i j} = W_{i \pi^{-1}(j)}$. Given that we include all possible permutations of our input configurations this immediately implies that all elements along the rows are constrained to be exactly identical ($W_{i j} = W_{i}$). Therefore, our weight matrix now has the form:

\begin{equation}
\mathbf{W} = \begin{bmatrix}
    \vec w_{1}\\
    \vdots \\
    \vec w_{K}
\end{bmatrix}, \textrm{where} \ \ \ \vec w_{i} \equiv W_{i} \underbrace{[1,1,\cdots,1]}_{L}.
\end{equation}

Enforcing the permutation symmetry effectively reduced the number of independent parameters from $L\times K$ in the general weight matrix to $K$ in the constrained case. This is reminiscent of the implementations of symmetry for translational invariance in \cite{5206577, sohn2012learning, Carleo_2017} where the weight matrix is reduced to a set of circulant matrices \cite{CIT-006} and each matrix can be treated as a convolutional kernel that is applied on all translated versions of the spin configuration $s$.  It is worth mentioning that permutation symmetry imposes a much stronger constraint than translation invariance.

As a key consequence of the above considerations, we can map the symmetry-imposed weight matrix to a new feed-forward network that takes as input the total magnetization $M_s  = \sum_{i} s_i$ and returns the value of $\log \Psi (s)$ (see Fig.\,\ref{fig:RBMilust}\textbf{B}). Using this structure, the symmetry-imposed output wave function can be written as:

\begin{equation}
\log \left( \langle s | \Psi \rangle \right) = \sum_{k=1}^{K} f(W_{k} M_s).
\label{eq:1paramanz}
\end{equation}

\noindent
It is important to note that this neural network wave function is not generally equivalent to a product state ansatz. Therefore, it opens the possibility to capture finite-size effects or correlations that could not be accounted for otherwise. Product states can still be captured by our ansatz as we will show in section \textbf{III}, where in the thermodynamic limit this ansatz becomes a product state asymptotically.

Therefore, when aiming to describe the ground state of a permutation-invariant Hamiltonian, Eq.\,$(6)$ provides an ansatz that accurately approximates the exact value for sufficiently many parameters $W_k$. This can be guaranteed due to the fact that multi-layer feed-forward neural networks have been shown to be universal function approximators \cite{HORNIK1989359}. Thus, for any targeted error $\epsilon$, there exists a $K$ such that 

\begin{equation}
\left \rvert \sum_{k=1}^{K} f(W_{k} M_s) - \log \Psi(s) \right \rvert < \epsilon.
\label{eq:bound}
\end{equation}

\noindent
That is, we may bound our approximation error by an arbitrary amount $\epsilon$ by tuning the value of $K$. The natural question is: How many parameters are necessary and how does this number depend on system size? We discuss this in detail in the remainder of the manuscript.

\section{\label{ssec:Finite-Size TFIM}Learning the fully-connected TFIM ground state}

We proceed by benchmarking the ansatz proposed in the previous section for a particular permutation-invariant Hamiltonian. This class of systems which give rise to mean-field models with permutation symmetry can be cast as long-range spin or boson models \cite{Sciolla_2013}. Concretely, we consider the fully-connected TFIM, whose Hamiltonian in terms of the Pauli matrices $S^{\rho}$\,($\rho = x,y,z$), takes the form:

\begin{equation}
    H = -\frac{J}{L} \sum_{i \neq j}^{} \ S^{z}_{i} S^{z}_{j} - g \sum_{i} S^{x}_{i}.
\end{equation}

This model exhibits a quantum phase transition at $g_c = 2J$ separating a ferromagnetic from a paramagnetic phase \cite{contextnote}. This critical point as well as ground states, and excited states have been studied previously in \cite{PhysRevB.74.144423,PhysRevB.104.085105,Sciolla_2013}. In the remainder of this work, we fix the value of $J = 1$. 

Although we expect that the choice of activation function is not crucial, we will take as ansatz the one defined in Eq.\,(\ref{eq:1paramanz}), with $f(x) = \log \left( \cosh(x) \right)$ as the activation function which is the choice that maps our model to a restricted Boltzmann machine which is a common reference for NQS \cite{Carleo_2017,PhysRevX.7.021021,PhysRevLett.125.100503}. Therefore, the wave function ansatz we will use in the following is given by:

\begin{equation}
     \log \Psi_{\mathbf{W}} (s) = \sum_{\ell=1}^{K} \log \cosh(W_{\ell} M_s).
\end{equation}

In order to numerically obtain the ground state of this model we will use stochastic reconfiguration \cite{PhysRevB.64.024512,PhysRevLett.80.4558} to minimize the variational energy $E(W)$ (see Eq.\,(\ref{eq:Ew})). It's of particular interest to study how the convergence of the model is affected by the choice of $K$. That is, for a given system size $L$ how many parameters do we need to converge to the exact ground state?

To evaluate the accuracy of this ansatz in finding the ground state of the fully-connected TFIM, we train the model and track both the relative energy:

\begin{equation}
    \epsilon_{\textrm{rel}} \equiv \frac{ |\langle H \rangle  - E_\textrm{ED}|}{ E_\textrm{ED}},
    \label{eq:relE}
\end{equation}
where the expectation values are defined as: $\langle \dots \rangle \equiv \langle \Psi_{\mathbf{W}}|\dots| \Psi_{\mathbf{W}} \rangle$, with appropriate normalization of the wave function and $E_\textrm{ED}$ denotes the ground state energy obtained by exact diagonalization.
As a further measure for the accuracy of our obtained wave function, we study also the energy fluctuation density defined as:
\begin{equation}
\sigma^2 (H) = \frac{1}{L} \Big[\langle H^2 \rangle - \langle H \rangle^2 \Big].
\label{eq:fluct}
\end{equation}
If our variational wave function is the targeted ground state, we would have $\sigma^2(H)=0$ on fundamental grounds.

In Fig.\,\ref{fig:RBMilust}\textbf{C} we show the results of the ground state search training for several choices of $K$ at a fixed small system size of $L=12$. The averages were computed by summing over the full Hilbert space basis configurations, which we refer to as exact sampling. We find that as one increases the number of independent variational parameters $K$, the model is able to lower the error further. This coincides with our claim emerging from the universal function approximator theorem in Eq.\,(\ref{eq:bound}), according to which we can reduce our error arbitrarily by taking more parameters $K$.

\begin{figure}[h!]
    \centering
    \includegraphics[width = 3.2in]{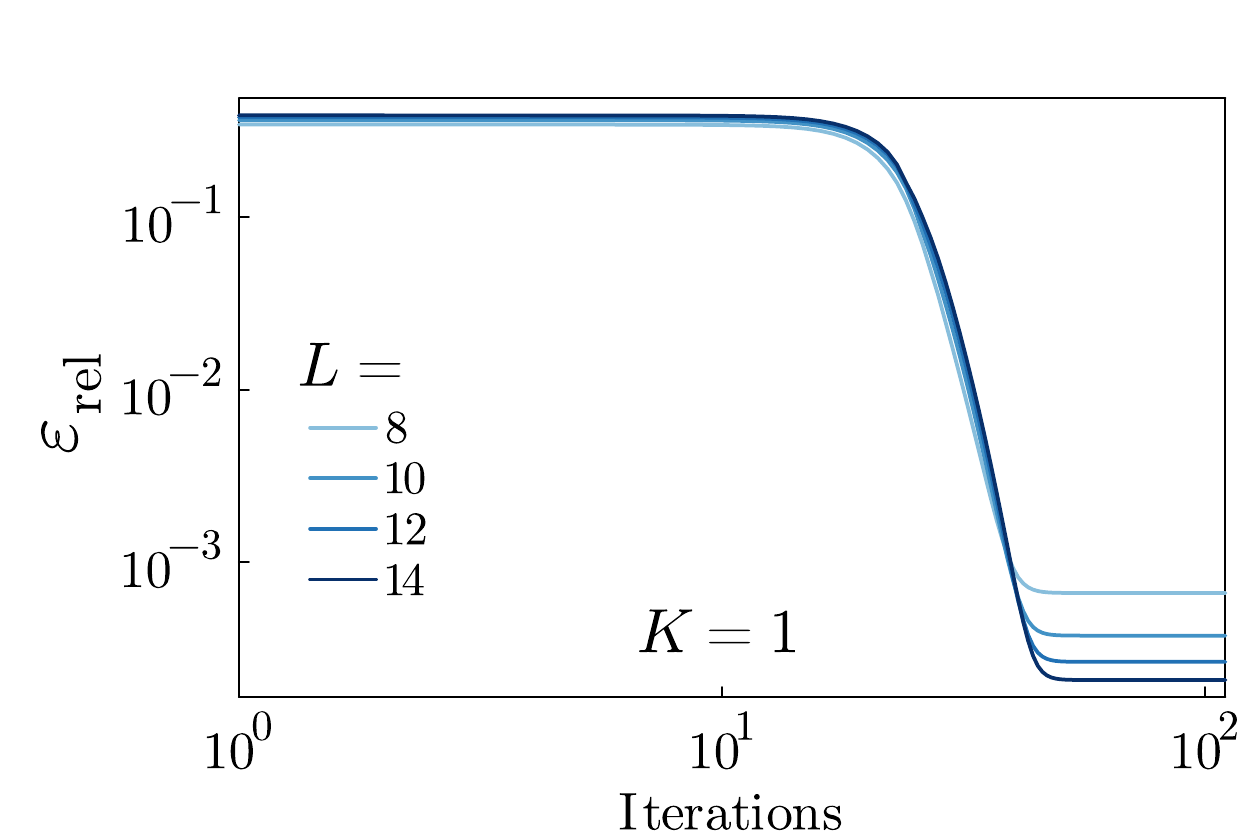}
    \caption{Relative energy for one variational parameter $K=1$ under training iterations for several system sizes $L$. We used exact sampling when averaging over spin configurations.}
    \label{fig:A0N1Erel}
\end{figure}

At this point, one further question arises: how do the parameter requirements scale with systems size? To address this question, in Fig.\,\ref{fig:A0N1Erel} we show the relative energy for several $L$ using only one parameter $K=1$. Here, we used exact diagonalization to compute the ground state energy and compare to our variational ansatz. It appears that as $L$ grows the $K=1$ result improves, signaling that convergence in fact requires fewer parameters as we increase system size. To further study this apparent behavior, in the following section, we will attempt to understand analytically how far we can get with a single parameter ansatz.

\subsection{Thermodynamic limit}

In this section, we demonstrate that the ground state of the fully-connected TFIM can be exactly described with only one variational parameter in the thermodynamic limit, building upon the observed trend in finite-size systems in Fig.\,\ref{fig:A0N1Erel}. We argue that this result signals the low complexity of permutation-symmetric models for neural quantum states. By taking the one-parameter case ($K=1$) we can take the following ansatz:

\begin{equation}
    \log \Psi_{W} (s) =  \log \left( \cosh(W M_s) \right).
\end{equation}

Due to the fact that this ansatz has only one variational parameter we can compute quantities such as the magnetization, energy, and the energy fluctuations analytically. This will allow us to show the asymptotic convergence to the ground state in the thermodynamic limit. To proceed with this computation we will restrict our system without loss of generality to the ferromagnetic phase and assume that $L$ is sufficiently large. Under these assumptions, the typical configurations $s$ are such that $|M_s| \sim L$, and hence we can approximate the $\log \cosh$ function as $\log (\cosh (WM_s)) \approx WM_s$ for $WM_s \sim L \gg 1$. We emphasize that the analytics conducted are applicable to any activation function exhibiting asymptotically linear behavior. This generality in our findings underscores the broader applicability of the results.

Details on all the following computations can be found in App.\,\ref{Ap: Analytic}. We first compute the energy

\begin{equation}
    E = \sum_{s}^{} E_{\textrm{loc}}(s) \frac{|\langle s | \Psi \rangle|^2}{\langle \Psi | \Psi \rangle}, \ \textrm{with} \ E_{\textrm{loc}}(s) \equiv \frac{\langle s |H|\Psi\rangle}{\langle s|\Psi\rangle}
    \label{eq:Ew}
\end{equation}

as a function of $W$ \cite{Carleo_2017}. For the energy we obtain:

\begin{multline}
    E(W) = -J(L-1) \tanh^{2}(2W) \\+ gL\tanh(2W) \sinh(2W) -gL\cosh(2W).
    \label{eq:EW}
\end{multline}

For the ground state we minimize Eq.\,(\ref{eq:EW}) yielding the value of $W$ in the ground state, which is given by:

\begin{equation}
W = \frac{1}{2} \cosh^{-1}\left[ \frac{2J}{g} \left(1 - \frac{1}{L}\right) \right].
\label{eq:WGS}
\end{equation}

\noindent
We can go one step further and relate this single parameter directly to the mean-field order parameter. If we compute the magnetization $M \equiv L^{-1} \sum_i \langle  S^{z}_i\rangle$, one finds that the parameter $W$ is related to the magnetization by

\begin{equation}
    M  = \tanh(2 W) = \pm \sqrt{1 - \frac{g^2}{4J^2}}.
    \label{eq:Magnetization}
\end{equation}

Where the second equality was obtained by substituting the value of $W$ in the ground state Eq.\,(\ref{eq:WGS}). It is worth mentioning that this equation is identical to the self-consistency equation for the mean-field solution of the nearest-neighbor Ising model at zero temperature, then we can identify that $W \sim M$. Hence, it serves as an example of how the network parameters can be directly related to physical quantities, and studying their structure may be beneficial to understand NQS complexity. As a next step, we show that this state is an eigenstate of our Hamiltonian asymptotically in $L$. To do this, we make use of the energy fluctuations. In App.\,\ref{Ap: Analytic}, we obtain as an analytical expression:

\begin{equation}
     \sigma^2 (H) = \frac{g^4 L^2}{8 J^2 (L-1)^3}.
     \label{eq:alp0sca}
\end{equation}

\noindent
This result indicates that the scale of the energy fluctuations asymptotically vanishes as $1/\sqrt{L}$. Implying that, in the thermodynamic limit, our ansatz becomes an exact eigenstate of the Hamiltonian. The one-parameter ansatz also provides plenty of information on the physics of the model. For instance, from Eq.\,(\ref{eq:WGS}) one may also extract the critical $g$ separating the ferromagnetic and paramagnetic region. This can be detected by noticing that the $\cosh^{-1}$ becomes undefined for $g>2J$. 

\section{\label{sec:LongRange}Breaking permutation symmetry}

We now study the effect of deviating from mean-field by breaking the permutation symmetry. For this, we will study the convergence to the ground state for the long-range interacting TFIM with periodic boundary conditions (PBC). The Hamiltonian for this model is given by:

\begin{equation}
    H = -\frac{J}{\mathcal{N}(L,\alpha)} \sum_{i \neq j}^{} \frac{1}{|i - j|^{\alpha}} \ S^{z}_{i} S^{z}_{j} - g \sum_{i} S^{x}_{i}.
\end{equation}

\noindent
Where $\mathcal{N}(L,\alpha) \equiv \frac{1}{L-1} \sum_{i\neq j}^{} \frac{1}{|i - j|^{\alpha}}$ is the so-called Kac normalization factor used to ensure that the energy is extensive. We take $|i-j|$ to be the minimum distance between two lattice sites in PBC. When $\alpha = 0$, the model is equivalent to the fully-connected TFIM studied before, while as $\alpha \rightarrow \infty$, the model reduces to the nearest-neighbor transverse-field Ising model. This model can be realized experimentally in trapped ion simulators for $0 \le \alpha \lesssim 3.5 $ \cite{Hauke_2013, Jurcevic_2014, 2008NatPh...4..757F, Islam_2011, schneider2011manybody,PRXQuantum.4.010302,PhysRevLett.92.207901} and is known to go through several phase transitions as a function of $\alpha$ \cite{PhysRevLett.109.267203,PhysRevB.64.184106,PhysRevB.64.184106, Dyson:1968up}. This setup is ideal for our purposes because we can adjust the deviation from mean-field upon tuning the value of $\alpha$, observing the eventual breakdown of our ansatz.

To verify the convergence to the ground state of this model as we vary the interaction range $\alpha$, we compute the energy fluctuation density $\sigma^2(H)$ of the one-parameter ansatz. We deduce an analytical expression which allows us to establish its system-size dependence:
\begin{multline}
   \sigma^2 (H) = g^2 \tanh^2(2W) -4Jg\left(1 - \frac{1}{L}\right) \frac{\tanh^2(2W)}{\cosh(2W)} \\ + \frac{4 J^2 (\tanh^2(2W)-\tanh^4(2W))}{ \mathcal{N}(L,\alpha)^2 \ L} \sum_{i\neq j \neq k} \frac{1}{|i-j|^{\alpha} |j-k|^{\alpha}} \\ + \frac{2 J^2  (1 - \tanh^4(2W))}{\mathcal{N}(L,\alpha)^2 \ L} \sum_{i\neq j} \frac{1}{|i-j|^{2\alpha}}.
   \label{eq:FluctAlp}
\end{multline}
\noindent
Details of this computation can be found in App.\,\ref{Ap: Analytic}. Given that we're interested in the thermodynamic limit behavior, we go ahead and rewrite the two remaining sums in terms of the generalized harmonic numbers $H_{n,r} \equiv \sum_{k=1}^{n} k^{-r} $. For $r > 1$, we make use of the fact that, in the limit when $n \rightarrow \infty$, $H_{n,r}$ converges to the Riemann zeta function defined as $\zeta(r) \equiv \sum_{k=1}^{\infty} k^{-r}$. This allows us to obtain the energy fluctuations in the thermodynamic limit directly as
\begin{equation}
\sigma^2 (H) = \frac{g^4}{16 J^2}
\begin{cases}
  \displaystyle\frac{H_{\infty,2 \alpha}}{H_{\infty, \alpha}^2}, & \text{if } \alpha \le 1 \\[13pt]
  \displaystyle\frac{\zeta(2 \alpha)}{\zeta(\alpha)^2}, & \text{if } \alpha > 1
\end{cases}.
\label{eq:FluctAlpRie}
\end{equation}

\noindent
By analyzing the generalized harmonic series ratio convergence as a function of $\alpha$ in  Eq.\,\,(\ref{eq:FluctAlpRie}), we find that in general, for $\alpha \le 1$ the energy fluctuations will vanish. This result implies that even though we have broken the permutation symmetry our ansatz is still able to exactly approach the ground state for large enough systems which is remarkable. 

\begin{figure}
    \centering
    \includegraphics[width = 3.2in]{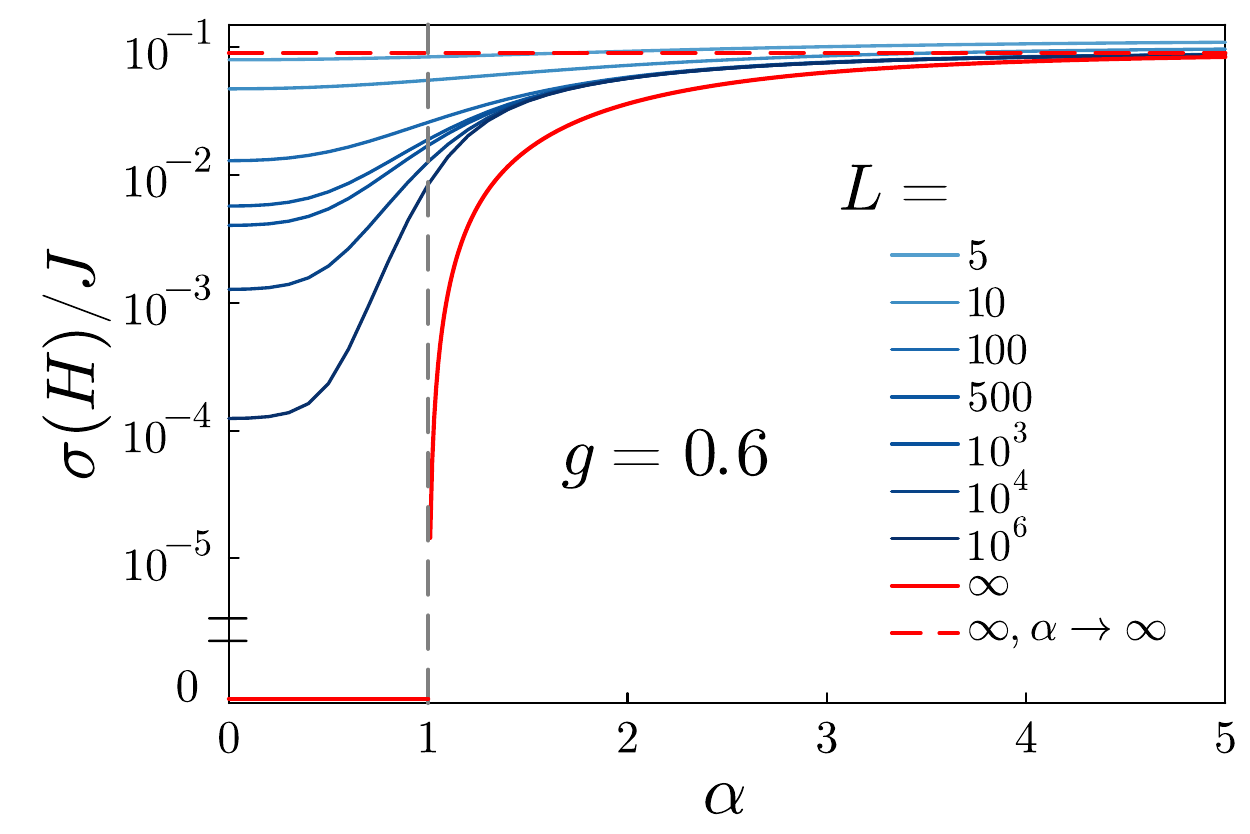}
    \caption{Normalized energy fluctuations of the long-range interacting transverse-field Ising model are shown as a function of interaction range $\alpha$ for various system sizes. The fluctuations are normalized by the coupling $J$ due to the inaccessibility of the gap in the thermodynamic limit for non-zero values of $\alpha$. In red we show the limiting curve in the thermodynamic limit.}
    \label{fig:FluctScaling}
\end{figure}

For the case of $\alpha > 1$, we no longer have diverging sums and can write the entire expression as a function of the ratio of the Riemann zeta functions. With this, we have the precise scaling with which the energy fluctuations decay as a function of $\alpha$ in the thermodynamic limit. In Fig.\,\ref{fig:FluctScaling}, we plot the expression $\sigma^2$ for different system sizes. Remarkably, we observe that the fluctuations vanish asymptotically as $1/\sqrt{L}$ for $\alpha \le 1/2$ precisely as for the $\alpha = 0$ mean-field case. In contrast, for $1/2 < \alpha \le 1$ the fluctuations still vanish asymptotically but with a slower decay. For this last interval the decays at the edges ($\alpha = 1/2,\, 1$) follow the scaling $\sqrt{\log(L)/L}$ and $\sqrt{\zeta(2)}/\log(L)$ respectively. For values of $\alpha > 1$, we also plot the limiting curve, where we see that the energy fluctuations steadily increase until saturating to an upper bound value of $g^2/4 J$. This value comes from evaluating the energy fluctuation density at the ground state with the $W$ given by Eq.\,(\ref{eq:WGS}).

The one-parameter ansatz consequently captures exactly the ground state for a finite range of alpha values. Interestingly, we find that the one-parameter ansatz can exactly describe quantum states even without permutation symmetry in the thermodynamic limit. Furthermore, it indicates that the long-range interacting TFIM has a simplified ground state within the range $0<\alpha\le 1$.

This type of long-range Ising model is known to exhibit quantum phase transitions as a function of $\alpha$, a Kosterlitz-Thouless transition at $\alpha = 1$, and for $\alpha \ge 2$ one recovers short-range behavior \cite{Dyson:1968up, Hauke_2013,PhysRevLett.109.267203,PhysRevB.64.184106}. However, the details about the ground-state properties in particular in the regime of low $\alpha$ have not been fully settled so far. For instance, in Ref.~\cite{PhysRevB.64.184106}, it has been found that for even larger $\alpha < 5/3$ the ground state might be described by a long-range mean-field theory. It is unclear, however, to which extent the utilized epsilon expansion in this high-$\alpha$ regime remains accurate.

For a comprehensive study of the prior findings related to this model, we would like to further direct the reader's attention to the recent review in Ref.\,\cite{defenu2023outofequilibrium}. It is essential to emphasize that while our conclusion of mean-field type behavior may appear intuitive, there has been a notable absence of formal proof establishing this models' ground state as a one-parameter simplified ansatz. While previous computations, employing linear spin wave theory for instance, have indeed yielded results consistent with our own observations, it is also noteworthy to emphasize that there has remained also a crucial difference. Within linear spin wave theory it has been shown that no inconsistencies are generated~\cite{defenu2023outofequilibrium}, which likely make the approach reliable, but still based on some unproven assumptions. Here, with our approach no such assumption is required providing us with a conclusive and rigorous proof.

\section{Conclusion}

In this work, we have investigated a class of models that demonstrate minimal complexity within the framework of NQS in the sense of the number of parameters required to describe the ground state. By imposing permutation symmetry on our networks, we have found that NQSs are particularly well-suited to describe permutation-invariant mean-field states. To validate this claim, we have investigated the ground state of the fully-connected TFIM. In particular, we show that,  even for finite systems, the ground state can be accurately captured using a small number of parameters, and this approximation improves as the system size increases.

An important advantage of our approach is the ability to capture finite-size effects, entanglement and quantum correlations, as it does not necessarily assume a product state. This is particularly relevant, as discussed in detail for instance in Ref.\,\cite{Sciolla_2013}, permutation symmetry does not always imply a product-state structure, especially for finite-sized systems or for dynamics. For the latter case it is also very important to note that such permutation-invariant Hamiltonians are key for the creation of spin squeezed states through one-axis twisting~\cite{Kitagawa1993}. Let us note, however, that we furthermore show that product states, such as those observed in the thermodynamic limit of our model, can still be achieved. Specifically, by opting for an activation function which behaves linearly at least asymptotically, we can witness the emergence of a product-state structure for the ground state in the thermodynamic limit.

Motivated by these observations, we have proposed a one-parameter ansatz and have analytically shown that the ground state can be described by a single variational parameter in the thermodynamic limit. Furthermore, we have examined the impact of breaking permutation symmetry by studying the ground state of the long-range interacting TFIM and tuning the interaction range. Surprisingly, we have found that the one-parameter ansatz
remains effective for values of $\alpha$ up to $1$, suggesting that the model is still described by mean-field in this regime and robustness in the presence of weak symmetry breaking.

A potential direction for future research lies in exploring the possibility of learning potential simplifications, such as the one discussed in this work for permutation symmetry, within the NQS parametrization directly from the network parameters. Such an approach may have the ability to shed light on the overall complexity of the network for more complicated scenarios. Additionally, an intriguing angle to consider is whether we can uncover underlying physics in the network parameters, such as the relation found here between the 1-parameter model and magnetization. In other words, can we demystify the black-box nature of the neural network and gain insights into the specific physical principles it exploits to represent the quantum state effectively?

While our work has shed some light on which quantum states are simple to capture with NQSs, it remains an open question to identify a general and physical characterization of states, which are difficult for NQSs. Recent developments have highlighted that even ground states of frustrated quantum magnets appear to be manageable within NQS~\cite{chen2023efficient}, although very deep networks seem to be necessary in order to be quantitatively accurate. This suggest that the intricate sign structure in such frustrated quantum magnets might be what is difficult to represent within NQSs. This would apply equivalently also to fermionic quantum matter. However, finally, this question will have to be explored in further detail in order to eventually arrive at a physical understanding of NQS complexity in the end.

\section*{Data availability}
The data shown in the figures is available on Zenodo \cite{ballar_trigueros_fabian_2023_8252947}.

\begin{acknowledgments}
We thank Marin Bukov, Alessio Lerose and Vighnesh Naik for fruitful discussions. Exact diagonalization was performed using QuSpin \cite{Weinberg_2017}. The NQS training was done leveraging the jVMC package \cite{Schmitt_2022} and the JAX library \cite{jax2018github}. This project has received funding from
the European Research Council (ERC) under the Euro-
pean Union’s Horizon 2020 research and innovation pro-
gramme (grant agreement No. 853443).
\end{acknowledgments}

\nocite{*}

\bibliography{NQSMFT}

\providecommand{\noopsort}[1]{}\providecommand{\singleletter}[1]{#1}%
\begin{thebibliography}{63}%
\makeatletter
\providecommand \@ifxundefined [1]{%
 \@ifx{#1\undefined}
}%
\providecommand \@ifnum [1]{%
 \ifnum #1\expandafter \@firstoftwo
 \else \expandafter \@secondoftwo
 \fi
}%
\providecommand \@ifx [1]{%
 \ifx #1\expandafter \@firstoftwo
 \else \expandafter \@secondoftwo
 \fi
}%
\providecommand \natexlab [1]{#1}%
\providecommand \enquote  [1]{``#1''}%
\providecommand \bibnamefont  [1]{#1}%
\providecommand \bibfnamefont [1]{#1}%
\providecommand \citenamefont [1]{#1}%
\providecommand \href@noop [0]{\@secondoftwo}%
\providecommand \href [0]{\begingroup \@sanitize@url \@href}%
\providecommand \@href[1]{\@@startlink{#1}\@@href}%
\providecommand \@@href[1]{\endgroup#1\@@endlink}%
\providecommand \@sanitize@url [0]{\catcode `\\12\catcode `\$12\catcode
  `\&12\catcode `\#12\catcode `\^12\catcode `\_12\catcode `\%12\relax}%
\providecommand \@@startlink[1]{}%
\providecommand \@@endlink[0]{}%
\providecommand \url  [0]{\begingroup\@sanitize@url \@url }%
\providecommand \@url [1]{\endgroup\@href {#1}{\urlprefix }}%
\providecommand \urlprefix  [0]{URL }%
\providecommand \Eprint [0]{\href }%
\providecommand \doibase [0]{https://doi.org/}%
\providecommand \selectlanguage [0]{\@gobble}%
\providecommand \bibinfo  [0]{\@secondoftwo}%
\providecommand \bibfield  [0]{\@secondoftwo}%
\providecommand \translation [1]{[#1]}%
\providecommand \BibitemOpen [0]{}%
\providecommand \bibitemStop [0]{}%
\providecommand \bibitemNoStop [0]{.\EOS\space}%
\providecommand \EOS [0]{\spacefactor3000\relax}%
\providecommand \BibitemShut  [1]{\csname bibitem#1\endcsname}%
\let\auto@bib@innerbib\@empty
\bibitem [{\citenamefont {Hohenberg}\ and\ \citenamefont
  {Kohn}(1964)}]{PhysRev.136.B864}%
  \BibitemOpen
  \bibfield  {author} {\bibinfo {author} {\bibfnamefont {P.}~\bibnamefont
  {Hohenberg}}\ and\ \bibinfo {author} {\bibfnamefont {W.}~\bibnamefont
  {Kohn}},\ }\bibfield  {title} {\bibinfo {title} {Inhomogeneous electron
  gas},\ }\href {https://doi.org/10.1103/PhysRev.136.B864} {\bibfield
  {journal} {\bibinfo  {journal} {Phys. Rev.}\ }\textbf {\bibinfo {volume}
  {136}},\ \bibinfo {pages} {B864} (\bibinfo {year} {1964})}\BibitemShut
  {NoStop}%
\bibitem [{\citenamefont {White}\ and\ \citenamefont
  {Feiguin}(2004)}]{PhysRevLett.93.076401}%
  \BibitemOpen
  \bibfield  {author} {\bibinfo {author} {\bibfnamefont {S.~R.}\ \bibnamefont
  {White}}\ and\ \bibinfo {author} {\bibfnamefont {A.~E.}\ \bibnamefont
  {Feiguin}},\ }\bibfield  {title} {\bibinfo {title} {Real-time evolution using
  the density matrix renormalization group},\ }\href
  {https://doi.org/10.1103/PhysRevLett.93.076401} {\bibfield  {journal}
  {\bibinfo  {journal} {Phys. Rev. Lett.}\ }\textbf {\bibinfo {volume} {93}},\
  \bibinfo {pages} {076401} (\bibinfo {year} {2004})}\BibitemShut {NoStop}%
\bibitem [{\citenamefont {Vidal}(2004)}]{PhysRevLett.93.040502}%
  \BibitemOpen
  \bibfield  {author} {\bibinfo {author} {\bibfnamefont {G.}~\bibnamefont
  {Vidal}},\ }\bibfield  {title} {\bibinfo {title} {Efficient simulation of
  one-dimensional quantum many-body systems},\ }\href
  {https://doi.org/10.1103/PhysRevLett.93.040502} {\bibfield  {journal}
  {\bibinfo  {journal} {Phys. Rev. Lett.}\ }\textbf {\bibinfo {volume} {93}},\
  \bibinfo {pages} {040502} (\bibinfo {year} {2004})}\BibitemShut {NoStop}%
\bibitem [{\citenamefont {White}(1992)}]{PhysRevLett.69.2863}%
  \BibitemOpen
  \bibfield  {author} {\bibinfo {author} {\bibfnamefont {S.~R.}\ \bibnamefont
  {White}},\ }\bibfield  {title} {\bibinfo {title} {Density matrix formulation
  for quantum renormalization groups},\ }\href
  {https://doi.org/10.1103/PhysRevLett.69.2863} {\bibfield  {journal} {\bibinfo
   {journal} {Phys. Rev. Lett.}\ }\textbf {\bibinfo {volume} {69}},\ \bibinfo
  {pages} {2863} (\bibinfo {year} {1992})}\BibitemShut {NoStop}%
\bibitem [{\citenamefont {Carlson}\ \emph {et~al.}(2015)\citenamefont
  {Carlson}, \citenamefont {Gandolfi}, \citenamefont {Pederiva}, \citenamefont
  {Pieper}, \citenamefont {Schiavilla}, \citenamefont {Schmidt},\ and\
  \citenamefont {Wiringa}}]{RevModPhys.87.1067}%
  \BibitemOpen
  \bibfield  {author} {\bibinfo {author} {\bibfnamefont {J.}~\bibnamefont
  {Carlson}}, \bibinfo {author} {\bibfnamefont {S.}~\bibnamefont {Gandolfi}},
  \bibinfo {author} {\bibfnamefont {F.}~\bibnamefont {Pederiva}}, \bibinfo
  {author} {\bibfnamefont {S.~C.}\ \bibnamefont {Pieper}}, \bibinfo {author}
  {\bibfnamefont {R.}~\bibnamefont {Schiavilla}}, \bibinfo {author}
  {\bibfnamefont {K.~E.}\ \bibnamefont {Schmidt}},\ and\ \bibinfo {author}
  {\bibfnamefont {R.~B.}\ \bibnamefont {Wiringa}},\ }\bibfield  {title}
  {\bibinfo {title} {Quantum monte carlo methods for nuclear physics},\ }\href
  {https://doi.org/10.1103/RevModPhys.87.1067} {\bibfield  {journal} {\bibinfo
  {journal} {Rev. Mod. Phys.}\ }\textbf {\bibinfo {volume} {87}},\ \bibinfo
  {pages} {1067} (\bibinfo {year} {2015})}\BibitemShut {NoStop}%
\bibitem [{\citenamefont {Foulkes}\ \emph {et~al.}(2001)\citenamefont
  {Foulkes}, \citenamefont {Mitas}, \citenamefont {Needs},\ and\ \citenamefont
  {Rajagopal}}]{RevModPhys.73.33}%
  \BibitemOpen
  \bibfield  {author} {\bibinfo {author} {\bibfnamefont {W.~M.~C.}\
  \bibnamefont {Foulkes}}, \bibinfo {author} {\bibfnamefont {L.}~\bibnamefont
  {Mitas}}, \bibinfo {author} {\bibfnamefont {R.~J.}\ \bibnamefont {Needs}},\
  and\ \bibinfo {author} {\bibfnamefont {G.}~\bibnamefont {Rajagopal}},\
  }\bibfield  {title} {\bibinfo {title} {Quantum monte carlo simulations of
  solids},\ }\href {https://doi.org/10.1103/RevModPhys.73.33} {\bibfield
  {journal} {\bibinfo  {journal} {Rev. Mod. Phys.}\ }\textbf {\bibinfo {volume}
  {73}},\ \bibinfo {pages} {33} (\bibinfo {year} {2001})}\BibitemShut {NoStop}%
\bibitem [{\citenamefont {Loh}\ \emph {et~al.}(1990)\citenamefont {Loh},
  \citenamefont {Gubernatis}, \citenamefont {Scalettar}, \citenamefont {White},
  \citenamefont {Scalapino},\ and\ \citenamefont {Sugar}}]{PhysRevB.41.9301}%
  \BibitemOpen
  \bibfield  {author} {\bibinfo {author} {\bibfnamefont {E.~Y.}\ \bibnamefont
  {Loh}}, \bibinfo {author} {\bibfnamefont {J.~E.}\ \bibnamefont {Gubernatis}},
  \bibinfo {author} {\bibfnamefont {R.~T.}\ \bibnamefont {Scalettar}}, \bibinfo
  {author} {\bibfnamefont {S.~R.}\ \bibnamefont {White}}, \bibinfo {author}
  {\bibfnamefont {D.~J.}\ \bibnamefont {Scalapino}},\ and\ \bibinfo {author}
  {\bibfnamefont {R.~L.}\ \bibnamefont {Sugar}},\ }\bibfield  {title} {\bibinfo
  {title} {Sign problem in the numerical simulation of many-electron systems},\
  }\href {https://doi.org/10.1103/PhysRevB.41.9301} {\bibfield  {journal}
  {\bibinfo  {journal} {Phys. Rev. B}\ }\textbf {\bibinfo {volume} {41}},\
  \bibinfo {pages} {9301} (\bibinfo {year} {1990})}\BibitemShut {NoStop}%
\bibitem [{\citenamefont {Perez-Garcia}\ \emph {et~al.}(2007)\citenamefont
  {Perez-Garcia}, \citenamefont {Verstraete}, \citenamefont {Wolf},\ and\
  \citenamefont {Cirac}}]{10.5555/2011832.2011833}%
  \BibitemOpen
  \bibfield  {author} {\bibinfo {author} {\bibfnamefont {D.}~\bibnamefont
  {Perez-Garcia}}, \bibinfo {author} {\bibfnamefont {F.}~\bibnamefont
  {Verstraete}}, \bibinfo {author} {\bibfnamefont {M.~M.}\ \bibnamefont
  {Wolf}},\ and\ \bibinfo {author} {\bibfnamefont {J.~I.}\ \bibnamefont
  {Cirac}},\ }\bibfield  {title} {\bibinfo {title} {Matrix product state
  representations},\ }\href@noop {} {\bibfield  {journal} {\bibinfo  {journal}
  {Quantum Info. Comput.}\ }\textbf {\bibinfo {volume} {7}},\ \bibinfo {pages}
  {401–430} (\bibinfo {year} {2007})}\BibitemShut {NoStop}%
\bibitem [{\citenamefont {Verstraete}\ and\ \citenamefont
  {Cirac}(2006)}]{PhysRevB.73.094423}%
  \BibitemOpen
  \bibfield  {author} {\bibinfo {author} {\bibfnamefont {F.}~\bibnamefont
  {Verstraete}}\ and\ \bibinfo {author} {\bibfnamefont {J.~I.}\ \bibnamefont
  {Cirac}},\ }\bibfield  {title} {\bibinfo {title} {Matrix product states
  represent ground states faithfully},\ }\href
  {https://doi.org/10.1103/PhysRevB.73.094423} {\bibfield  {journal} {\bibinfo
  {journal} {Phys. Rev. B}\ }\textbf {\bibinfo {volume} {73}},\ \bibinfo
  {pages} {094423} (\bibinfo {year} {2006})}\BibitemShut {NoStop}%
\bibitem [{\citenamefont {Carleo}\ and\ \citenamefont
  {Troyer}(2017)}]{Carleo_2017}%
  \BibitemOpen
  \bibfield  {author} {\bibinfo {author} {\bibfnamefont {G.}~\bibnamefont
  {Carleo}}\ and\ \bibinfo {author} {\bibfnamefont {M.}~\bibnamefont
  {Troyer}},\ }\bibfield  {title} {\bibinfo {title} {Solving the quantum
  many-body problem with artificial neural networks},\ }\href
  {https://doi.org/10.1126/science.aag2302} {\bibfield  {journal} {\bibinfo
  {journal} {Science}\ }\textbf {\bibinfo {volume} {355}},\ \bibinfo {pages}
  {602} (\bibinfo {year} {2017})}\BibitemShut {NoStop}%
\bibitem [{\citenamefont {Schmitt}\ and\ \citenamefont
  {Heyl}(2020)}]{PhysRevLett.125.100503}%
  \BibitemOpen
  \bibfield  {author} {\bibinfo {author} {\bibfnamefont {M.}~\bibnamefont
  {Schmitt}}\ and\ \bibinfo {author} {\bibfnamefont {M.}~\bibnamefont {Heyl}},\
  }\bibfield  {title} {\bibinfo {title} {Quantum many-body dynamics in two
  dimensions with artificial neural networks},\ }\href
  {https://doi.org/10.1103/PhysRevLett.125.100503} {\bibfield  {journal}
  {\bibinfo  {journal} {Phys. Rev. Lett.}\ }\textbf {\bibinfo {volume} {125}},\
  \bibinfo {pages} {100503} (\bibinfo {year} {2020})}\BibitemShut {NoStop}%
\bibitem [{\citenamefont {Chen}\ and\ \citenamefont
  {Heyl}(2023)}]{chen2023efficient}%
  \BibitemOpen
  \bibfield  {author} {\bibinfo {author} {\bibfnamefont {A.}~\bibnamefont
  {Chen}}\ and\ \bibinfo {author} {\bibfnamefont {M.}~\bibnamefont {Heyl}},\
  }\href@noop {} {\bibinfo {title} {Efficient optimization of deep neural
  quantum states toward machine precision}} (\bibinfo {year} {2023}),\ \Eprint
  {https://arxiv.org/abs/2302.01941} {arXiv:2302.01941 [cond-mat.dis-nn]}
  \BibitemShut {NoStop}%
\bibitem [{\citenamefont {Mendes-Santos}\ \emph {et~al.}(2023)\citenamefont
  {Mendes-Santos}, \citenamefont {Schmitt},\ and\ \citenamefont
  {Heyl}}]{Mendes_Santos_2023}%
  \BibitemOpen
  \bibfield  {author} {\bibinfo {author} {\bibfnamefont {T.}~\bibnamefont
  {Mendes-Santos}}, \bibinfo {author} {\bibfnamefont {M.}~\bibnamefont
  {Schmitt}},\ and\ \bibinfo {author} {\bibfnamefont {M.}~\bibnamefont
  {Heyl}},\ }\bibfield  {title} {\bibinfo {title} {Highly resolved spectral
  functions of two-dimensional systems with neural quantum states},\ }\bibfield
   {journal} {\bibinfo  {journal} {Physical Review Letters}\ }\textbf {\bibinfo
  {volume} {131}},\ \href {https://doi.org/10.1103/physrevlett.131.046501}
  {10.1103/physrevlett.131.046501} (\bibinfo {year} {2023})\BibitemShut
  {NoStop}%
\bibitem [{\citenamefont {Moreno}\ \emph {et~al.}(2022)\citenamefont {Moreno},
  \citenamefont {Carleo}, \citenamefont {Georges},\ and\ \citenamefont
  {Stokes}}]{doi:10.1073/pnas.2122059119}%
  \BibitemOpen
  \bibfield  {author} {\bibinfo {author} {\bibfnamefont {J.~R.}\ \bibnamefont
  {Moreno}}, \bibinfo {author} {\bibfnamefont {G.}~\bibnamefont {Carleo}},
  \bibinfo {author} {\bibfnamefont {A.}~\bibnamefont {Georges}},\ and\ \bibinfo
  {author} {\bibfnamefont {J.}~\bibnamefont {Stokes}},\ }\bibfield  {title}
  {\bibinfo {title} {Fermionic wave functions from neural-network constrained
  hidden states},\ }\href {https://doi.org/10.1073/pnas.2122059119} {\bibfield
  {journal} {\bibinfo  {journal} {Proceedings of the National Academy of
  Sciences}\ }\textbf {\bibinfo {volume} {119}},\ \bibinfo {pages}
  {e2122059119} (\bibinfo {year} {2022})},\ \Eprint
  {https://arxiv.org/abs/https://www.pnas.org/doi/pdf/10.1073/pnas.2122059119}
  {https://www.pnas.org/doi/pdf/10.1073/pnas.2122059119} \BibitemShut {NoStop}%
\bibitem [{\citenamefont {Choo}\ \emph {et~al.}(2019)\citenamefont {Choo},
  \citenamefont {Neupert},\ and\ \citenamefont {Carleo}}]{PhysRevB.100.125124}%
  \BibitemOpen
  \bibfield  {author} {\bibinfo {author} {\bibfnamefont {K.}~\bibnamefont
  {Choo}}, \bibinfo {author} {\bibfnamefont {T.}~\bibnamefont {Neupert}},\ and\
  \bibinfo {author} {\bibfnamefont {G.}~\bibnamefont {Carleo}},\ }\bibfield
  {title} {\bibinfo {title} {Two-dimensional frustrated
  ${J}_{1}\text{\ensuremath{-}}{J}_{2}$ model studied with neural network
  quantum states},\ }\href {https://doi.org/10.1103/PhysRevB.100.125124}
  {\bibfield  {journal} {\bibinfo  {journal} {Phys. Rev. B}\ }\textbf {\bibinfo
  {volume} {100}},\ \bibinfo {pages} {125124} (\bibinfo {year}
  {2019})}\BibitemShut {NoStop}%
\bibitem [{\citenamefont {Deng}\ \emph {et~al.}(2017)\citenamefont {Deng},
  \citenamefont {Li},\ and\ \citenamefont {Das~Sarma}}]{PhysRevX.7.021021}%
  \BibitemOpen
  \bibfield  {author} {\bibinfo {author} {\bibfnamefont {D.-L.}\ \bibnamefont
  {Deng}}, \bibinfo {author} {\bibfnamefont {X.}~\bibnamefont {Li}},\ and\
  \bibinfo {author} {\bibfnamefont {S.}~\bibnamefont {Das~Sarma}},\ }\bibfield
  {title} {\bibinfo {title} {Quantum entanglement in neural network states},\
  }\href {https://doi.org/10.1103/PhysRevX.7.021021} {\bibfield  {journal}
  {\bibinfo  {journal} {Phys. Rev. X}\ }\textbf {\bibinfo {volume} {7}},\
  \bibinfo {pages} {021021} (\bibinfo {year} {2017})}\BibitemShut {NoStop}%
\bibitem [{\citenamefont {Carleo}\ \emph {et~al.}(2018)\citenamefont {Carleo},
  \citenamefont {Nomura},\ and\ \citenamefont {Imada}}]{Carleo_2018}%
  \BibitemOpen
  \bibfield  {author} {\bibinfo {author} {\bibfnamefont {G.}~\bibnamefont
  {Carleo}}, \bibinfo {author} {\bibfnamefont {Y.}~\bibnamefont {Nomura}},\
  and\ \bibinfo {author} {\bibfnamefont {M.}~\bibnamefont {Imada}},\ }\bibfield
   {title} {\bibinfo {title} {Constructing exact representations of quantum
  many-body systems with deep neural networks},\ }\bibfield  {journal}
  {\bibinfo  {journal} {Nature Communications}\ }\textbf {\bibinfo {volume}
  {9}},\ \href {https://doi.org/10.1038/s41467-018-07520-3}
  {10.1038/s41467-018-07520-3} (\bibinfo {year} {2018})\BibitemShut {NoStop}%
\bibitem [{\citenamefont {Sharir}\ \emph {et~al.}(2022)\citenamefont {Sharir},
  \citenamefont {Shashua},\ and\ \citenamefont {Carleo}}]{PhysRevB.106.205136}%
  \BibitemOpen
  \bibfield  {author} {\bibinfo {author} {\bibfnamefont {O.}~\bibnamefont
  {Sharir}}, \bibinfo {author} {\bibfnamefont {A.}~\bibnamefont {Shashua}},\
  and\ \bibinfo {author} {\bibfnamefont {G.}~\bibnamefont {Carleo}},\
  }\bibfield  {title} {\bibinfo {title} {Neural tensor contractions and the
  expressive power of deep neural quantum states},\ }\href
  {https://doi.org/10.1103/PhysRevB.106.205136} {\bibfield  {journal} {\bibinfo
   {journal} {Phys. Rev. B}\ }\textbf {\bibinfo {volume} {106}},\ \bibinfo
  {pages} {205136} (\bibinfo {year} {2022})}\BibitemShut {NoStop}%
\bibitem [{\citenamefont {Norouzi}\ \emph {et~al.}(2009)\citenamefont
  {Norouzi}, \citenamefont {Ranjbar},\ and\ \citenamefont {Mori}}]{5206577}%
  \BibitemOpen
  \bibfield  {author} {\bibinfo {author} {\bibfnamefont {M.}~\bibnamefont
  {Norouzi}}, \bibinfo {author} {\bibfnamefont {M.}~\bibnamefont {Ranjbar}},\
  and\ \bibinfo {author} {\bibfnamefont {G.}~\bibnamefont {Mori}},\ }\bibfield
  {title} {\bibinfo {title} {Stacks of convolutional restricted boltzmann
  machines for shift-invariant feature learning},\ }in\ \href
  {https://doi.org/10.1109/CVPR.2009.5206577} {\emph {\bibinfo {booktitle}
  {2009 IEEE Conference on Computer Vision and Pattern Recognition}}}\
  (\bibinfo {year} {2009})\ pp.\ \bibinfo {pages} {2735--2742}\BibitemShut
  {NoStop}%
\bibitem [{\citenamefont {Sohn}\ and\ \citenamefont
  {Lee}(2012)}]{sohn2012learning}%
  \BibitemOpen
  \bibfield  {author} {\bibinfo {author} {\bibfnamefont {K.}~\bibnamefont
  {Sohn}}\ and\ \bibinfo {author} {\bibfnamefont {H.}~\bibnamefont {Lee}},\
  }\href@noop {} {\bibinfo {title} {Learning invariant representations with
  local transformations}} (\bibinfo {year} {2012}),\ \Eprint
  {https://arxiv.org/abs/1206.6418} {arXiv:1206.6418 [cs.LG]} \BibitemShut
  {NoStop}%
\bibitem [{\citenamefont {Gray}(2006)}]{CIT-006}%
  \BibitemOpen
  \bibfield  {author} {\bibinfo {author} {\bibfnamefont {R.~M.}\ \bibnamefont
  {Gray}},\ }\bibfield  {title} {\bibinfo {title} {Toeplitz and circulant
  matrices: A review},\ }\href {https://doi.org/10.1561/0100000006} {\bibfield
  {journal} {\bibinfo  {journal} {Foundations and Trends® in Communications
  and Information Theory}\ }\textbf {\bibinfo {volume} {2}},\ \bibinfo {pages}
  {155} (\bibinfo {year} {2006})}\BibitemShut {NoStop}%
\bibitem [{\citenamefont {Hornik}\ \emph {et~al.}(1989)\citenamefont {Hornik},
  \citenamefont {Stinchcombe},\ and\ \citenamefont {White}}]{HORNIK1989359}%
  \BibitemOpen
  \bibfield  {author} {\bibinfo {author} {\bibfnamefont {K.}~\bibnamefont
  {Hornik}}, \bibinfo {author} {\bibfnamefont {M.}~\bibnamefont
  {Stinchcombe}},\ and\ \bibinfo {author} {\bibfnamefont {H.}~\bibnamefont
  {White}},\ }\bibfield  {title} {\bibinfo {title} {Multilayer feedforward
  networks are universal approximators},\ }\href
  {https://doi.org/https://doi.org/10.1016/0893-6080(89)90020-8} {\bibfield
  {journal} {\bibinfo  {journal} {Neural Networks}\ }\textbf {\bibinfo {volume}
  {2}},\ \bibinfo {pages} {359} (\bibinfo {year} {1989})}\BibitemShut {NoStop}%
\bibitem [{\citenamefont {Sciolla}\ and\ \citenamefont
  {Biroli}(2013)}]{Sciolla_2013}%
  \BibitemOpen
  \bibfield  {author} {\bibinfo {author} {\bibfnamefont {B.}~\bibnamefont
  {Sciolla}}\ and\ \bibinfo {author} {\bibfnamefont {G.}~\bibnamefont
  {Biroli}},\ }\bibfield  {title} {\bibinfo {title} {Quantum quenches,
  dynamical transitions, and off-equilibrium quantum criticality},\ }\bibfield
  {journal} {\bibinfo  {journal} {Physical Review B}\ }\textbf {\bibinfo
  {volume} {88}},\ \href {https://doi.org/10.1103/physrevb.88.201110}
  {10.1103/physrevb.88.201110} (\bibinfo {year} {2013})\BibitemShut {NoStop}%
\bibitem [{con()}]{contextnote}%
  \BibitemOpen
  \href@noop {} {\bibinfo {title} {Due to our definition of the hamiltonian the
  critical parameter ratio looks like this. {I}n other conventions it is
  reported as ${J = g}$ which is equivalent.}}\BibitemShut {Stop}%
\bibitem [{\citenamefont {Das}\ \emph {et~al.}(2006)\citenamefont {Das},
  \citenamefont {Sengupta}, \citenamefont {Sen},\ and\ \citenamefont
  {Chakrabarti}}]{PhysRevB.74.144423}%
  \BibitemOpen
  \bibfield  {author} {\bibinfo {author} {\bibfnamefont {A.}~\bibnamefont
  {Das}}, \bibinfo {author} {\bibfnamefont {K.}~\bibnamefont {Sengupta}},
  \bibinfo {author} {\bibfnamefont {D.}~\bibnamefont {Sen}},\ and\ \bibinfo
  {author} {\bibfnamefont {B.~K.}\ \bibnamefont {Chakrabarti}},\ }\bibfield
  {title} {\bibinfo {title} {Infinite-range ising ferromagnet in a
  time-dependent transverse magnetic field: Quench and ac dynamics near the
  quantum critical point},\ }\href {https://doi.org/10.1103/PhysRevB.74.144423}
  {\bibfield  {journal} {\bibinfo  {journal} {Phys. Rev. B}\ }\textbf {\bibinfo
  {volume} {74}},\ \bibinfo {pages} {144423} (\bibinfo {year}
  {2006})}\BibitemShut {NoStop}%
\bibitem [{\citenamefont {Sehrawat}\ \emph {et~al.}(2021)\citenamefont
  {Sehrawat}, \citenamefont {Srivastava},\ and\ \citenamefont
  {Sen}}]{PhysRevB.104.085105}%
  \BibitemOpen
  \bibfield  {author} {\bibinfo {author} {\bibfnamefont {A.}~\bibnamefont
  {Sehrawat}}, \bibinfo {author} {\bibfnamefont {C.}~\bibnamefont
  {Srivastava}},\ and\ \bibinfo {author} {\bibfnamefont {U.}~\bibnamefont
  {Sen}},\ }\bibfield  {title} {\bibinfo {title} {Dynamical phase transitions
  in the fully connected quantum ising model: Time period and critical time},\
  }\href {https://doi.org/10.1103/PhysRevB.104.085105} {\bibfield  {journal}
  {\bibinfo  {journal} {Phys. Rev. B}\ }\textbf {\bibinfo {volume} {104}},\
  \bibinfo {pages} {085105} (\bibinfo {year} {2021})}\BibitemShut {NoStop}%
\bibitem [{\citenamefont {Sorella}(2001)}]{PhysRevB.64.024512}%
  \BibitemOpen
  \bibfield  {author} {\bibinfo {author} {\bibfnamefont {S.}~\bibnamefont
  {Sorella}},\ }\bibfield  {title} {\bibinfo {title} {Generalized lanczos
  algorithm for variational quantum monte carlo},\ }\href
  {https://doi.org/10.1103/PhysRevB.64.024512} {\bibfield  {journal} {\bibinfo
  {journal} {Phys. Rev. B}\ }\textbf {\bibinfo {volume} {64}},\ \bibinfo
  {pages} {024512} (\bibinfo {year} {2001})}\BibitemShut {NoStop}%
\bibitem [{\citenamefont {Sorella}(1998)}]{PhysRevLett.80.4558}%
  \BibitemOpen
  \bibfield  {author} {\bibinfo {author} {\bibfnamefont {S.}~\bibnamefont
  {Sorella}},\ }\bibfield  {title} {\bibinfo {title} {Green function monte
  carlo with stochastic reconfiguration},\ }\href
  {https://doi.org/10.1103/PhysRevLett.80.4558} {\bibfield  {journal} {\bibinfo
   {journal} {Phys. Rev. Lett.}\ }\textbf {\bibinfo {volume} {80}},\ \bibinfo
  {pages} {4558} (\bibinfo {year} {1998})}\BibitemShut {NoStop}%
\bibitem [{\citenamefont {Hauke}\ and\ \citenamefont
  {Tagliacozzo}(2013)}]{Hauke_2013}%
  \BibitemOpen
  \bibfield  {author} {\bibinfo {author} {\bibfnamefont {P.}~\bibnamefont
  {Hauke}}\ and\ \bibinfo {author} {\bibfnamefont {L.}~\bibnamefont
  {Tagliacozzo}},\ }\bibfield  {title} {\bibinfo {title} {Spread of
  correlations in long-range interacting quantum systems},\ }\bibfield
  {journal} {\bibinfo  {journal} {Physical Review Letters}\ }\textbf {\bibinfo
  {volume} {111}},\ \href {https://doi.org/10.1103/physrevlett.111.207202}
  {10.1103/physrevlett.111.207202} (\bibinfo {year} {2013})\BibitemShut
  {NoStop}%
\bibitem [{\citenamefont {Jurcevic}\ \emph {et~al.}(2014)\citenamefont
  {Jurcevic}, \citenamefont {Lanyon}, \citenamefont {Hauke}, \citenamefont
  {Hempel}, \citenamefont {Zoller}, \citenamefont {Blatt},\ and\ \citenamefont
  {Roos}}]{Jurcevic_2014}%
  \BibitemOpen
  \bibfield  {author} {\bibinfo {author} {\bibfnamefont {P.}~\bibnamefont
  {Jurcevic}}, \bibinfo {author} {\bibfnamefont {B.~P.}\ \bibnamefont
  {Lanyon}}, \bibinfo {author} {\bibfnamefont {P.}~\bibnamefont {Hauke}},
  \bibinfo {author} {\bibfnamefont {C.}~\bibnamefont {Hempel}}, \bibinfo
  {author} {\bibfnamefont {P.}~\bibnamefont {Zoller}}, \bibinfo {author}
  {\bibfnamefont {R.}~\bibnamefont {Blatt}},\ and\ \bibinfo {author}
  {\bibfnamefont {C.~F.}\ \bibnamefont {Roos}},\ }\bibfield  {title} {\bibinfo
  {title} {Quasiparticle engineering and entanglement propagation in a quantum
  many-body system},\ }\href {https://doi.org/10.1038/nature13461} {\bibfield
  {journal} {\bibinfo  {journal} {Nature}\ }\textbf {\bibinfo {volume} {511}},\
  \bibinfo {pages} {202} (\bibinfo {year} {2014})}\BibitemShut {NoStop}%
\bibitem [{\citenamefont {{Friedenauer}}\ \emph {et~al.}(2008)\citenamefont
  {{Friedenauer}}, \citenamefont {{Schmitz}}, \citenamefont {{Glueckert}},
  \citenamefont {{Porras}},\ and\ \citenamefont
  {{Schaetz}}}]{2008NatPh...4..757F}%
  \BibitemOpen
  \bibfield  {author} {\bibinfo {author} {\bibfnamefont {A.}~\bibnamefont
  {{Friedenauer}}}, \bibinfo {author} {\bibfnamefont {H.}~\bibnamefont
  {{Schmitz}}}, \bibinfo {author} {\bibfnamefont {J.~T.}\ \bibnamefont
  {{Glueckert}}}, \bibinfo {author} {\bibfnamefont {D.}~\bibnamefont
  {{Porras}}},\ and\ \bibinfo {author} {\bibfnamefont {T.}~\bibnamefont
  {{Schaetz}}},\ }\bibfield  {title} {\bibinfo {title} {{Simulating a quantum
  magnet with trapped ions}},\ }\href {https://doi.org/10.1038/nphys1032}
  {\bibfield  {journal} {\bibinfo  {journal} {Nature Physics}\ }\textbf
  {\bibinfo {volume} {4}},\ \bibinfo {pages} {757} (\bibinfo {year}
  {2008})}\BibitemShut {NoStop}%
\bibitem [{\citenamefont {Islam}\ \emph {et~al.}(2011)\citenamefont {Islam},
  \citenamefont {Edwards}, \citenamefont {Kim}, \citenamefont {Korenblit},
  \citenamefont {Noh}, \citenamefont {Carmichael}, \citenamefont {Lin},
  \citenamefont {Duan}, \citenamefont {Wang}, \citenamefont {Freericks},\ and\
  \citenamefont {Monroe}}]{Islam_2011}%
  \BibitemOpen
  \bibfield  {author} {\bibinfo {author} {\bibfnamefont {R.}~\bibnamefont
  {Islam}}, \bibinfo {author} {\bibfnamefont {E.}~\bibnamefont {Edwards}},
  \bibinfo {author} {\bibfnamefont {K.}~\bibnamefont {Kim}}, \bibinfo {author}
  {\bibfnamefont {S.}~\bibnamefont {Korenblit}}, \bibinfo {author}
  {\bibfnamefont {C.}~\bibnamefont {Noh}}, \bibinfo {author} {\bibfnamefont
  {H.}~\bibnamefont {Carmichael}}, \bibinfo {author} {\bibfnamefont {G.-D.}\
  \bibnamefont {Lin}}, \bibinfo {author} {\bibfnamefont {L.-M.}\ \bibnamefont
  {Duan}}, \bibinfo {author} {\bibfnamefont {C.-C.~J.}\ \bibnamefont {Wang}},
  \bibinfo {author} {\bibfnamefont {J.}~\bibnamefont {Freericks}},\ and\
  \bibinfo {author} {\bibfnamefont {C.}~\bibnamefont {Monroe}},\ }\bibfield
  {title} {\bibinfo {title} {Onset of a quantum phase transition with a trapped
  ion quantum simulator},\ }\bibfield  {journal} {\bibinfo  {journal} {Nature
  Communications}\ }\textbf {\bibinfo {volume} {2}},\ \href
  {https://doi.org/10.1038/ncomms1374} {10.1038/ncomms1374} (\bibinfo {year}
  {2011})\BibitemShut {NoStop}%
\bibitem [{\citenamefont {Schneider}\ \emph {et~al.}(2011)\citenamefont
  {Schneider}, \citenamefont {Porras},\ and\ \citenamefont
  {Schaetz}}]{schneider2011manybody}%
  \BibitemOpen
  \bibfield  {author} {\bibinfo {author} {\bibfnamefont {C.}~\bibnamefont
  {Schneider}}, \bibinfo {author} {\bibfnamefont {D.}~\bibnamefont {Porras}},\
  and\ \bibinfo {author} {\bibfnamefont {T.}~\bibnamefont {Schaetz}},\
  }\href@noop {} {\bibinfo {title} {Many-body physics with trapped ions}}
  (\bibinfo {year} {2011}),\ \Eprint {https://arxiv.org/abs/1106.2597}
  {arXiv:1106.2597 [quant-ph]} \BibitemShut {NoStop}%
\bibitem [{\citenamefont {Li}\ \emph {et~al.}(2023)\citenamefont {Li},
  \citenamefont {Wu}, \citenamefont {Mei}, \citenamefont {Yao}, \citenamefont
  {Lian}, \citenamefont {Cai}, \citenamefont {Wang}, \citenamefont {Qi},
  \citenamefont {Yao}, \citenamefont {He}, \citenamefont {Zhou},\ and\
  \citenamefont {Duan}}]{PRXQuantum.4.010302}%
  \BibitemOpen
  \bibfield  {author} {\bibinfo {author} {\bibfnamefont {B.-W.}\ \bibnamefont
  {Li}}, \bibinfo {author} {\bibfnamefont {Y.-K.}\ \bibnamefont {Wu}}, \bibinfo
  {author} {\bibfnamefont {Q.-X.}\ \bibnamefont {Mei}}, \bibinfo {author}
  {\bibfnamefont {R.}~\bibnamefont {Yao}}, \bibinfo {author} {\bibfnamefont
  {W.-Q.}\ \bibnamefont {Lian}}, \bibinfo {author} {\bibfnamefont {M.-L.}\
  \bibnamefont {Cai}}, \bibinfo {author} {\bibfnamefont {Y.}~\bibnamefont
  {Wang}}, \bibinfo {author} {\bibfnamefont {B.-X.}\ \bibnamefont {Qi}},
  \bibinfo {author} {\bibfnamefont {L.}~\bibnamefont {Yao}}, \bibinfo {author}
  {\bibfnamefont {L.}~\bibnamefont {He}}, \bibinfo {author} {\bibfnamefont
  {Z.-C.}\ \bibnamefont {Zhou}},\ and\ \bibinfo {author} {\bibfnamefont
  {L.-M.}\ \bibnamefont {Duan}},\ }\bibfield  {title} {\bibinfo {title}
  {Probing critical behavior of long-range transverse-field ising model through
  quantum kibble-zurek mechanism},\ }\href
  {https://doi.org/10.1103/PRXQuantum.4.010302} {\bibfield  {journal} {\bibinfo
   {journal} {PRX Quantum}\ }\textbf {\bibinfo {volume} {4}},\ \bibinfo {pages}
  {010302} (\bibinfo {year} {2023})}\BibitemShut {NoStop}%
\bibitem [{\citenamefont {Porras}\ and\ \citenamefont
  {Cirac}(2004)}]{PhysRevLett.92.207901}%
  \BibitemOpen
  \bibfield  {author} {\bibinfo {author} {\bibfnamefont {D.}~\bibnamefont
  {Porras}}\ and\ \bibinfo {author} {\bibfnamefont {J.~I.}\ \bibnamefont
  {Cirac}},\ }\bibfield  {title} {\bibinfo {title} {Effective quantum spin
  systems with trapped ions},\ }\href
  {https://doi.org/10.1103/PhysRevLett.92.207901} {\bibfield  {journal}
  {\bibinfo  {journal} {Phys. Rev. Lett.}\ }\textbf {\bibinfo {volume} {92}},\
  \bibinfo {pages} {207901} (\bibinfo {year} {2004})}\BibitemShut {NoStop}%
\bibitem [{\citenamefont {Koffel}\ \emph {et~al.}(2012)\citenamefont {Koffel},
  \citenamefont {Lewenstein},\ and\ \citenamefont
  {Tagliacozzo}}]{PhysRevLett.109.267203}%
  \BibitemOpen
  \bibfield  {author} {\bibinfo {author} {\bibfnamefont {T.}~\bibnamefont
  {Koffel}}, \bibinfo {author} {\bibfnamefont {M.}~\bibnamefont {Lewenstein}},\
  and\ \bibinfo {author} {\bibfnamefont {L.}~\bibnamefont {Tagliacozzo}},\
  }\bibfield  {title} {\bibinfo {title} {Entanglement entropy for the
  long-range ising chain in a transverse field},\ }\href
  {https://doi.org/10.1103/PhysRevLett.109.267203} {\bibfield  {journal}
  {\bibinfo  {journal} {Phys. Rev. Lett.}\ }\textbf {\bibinfo {volume} {109}},\
  \bibinfo {pages} {267203} (\bibinfo {year} {2012})}\BibitemShut {NoStop}%
\bibitem [{\citenamefont {Dutta}\ and\ \citenamefont
  {Bhattacharjee}(2001)}]{PhysRevB.64.184106}%
  \BibitemOpen
  \bibfield  {author} {\bibinfo {author} {\bibfnamefont {A.}~\bibnamefont
  {Dutta}}\ and\ \bibinfo {author} {\bibfnamefont {J.~K.}\ \bibnamefont
  {Bhattacharjee}},\ }\bibfield  {title} {\bibinfo {title} {Phase transitions
  in the quantum ising and rotor models with a long-range interaction},\ }\href
  {https://doi.org/10.1103/PhysRevB.64.184106} {\bibfield  {journal} {\bibinfo
  {journal} {Phys. Rev. B}\ }\textbf {\bibinfo {volume} {64}},\ \bibinfo
  {pages} {184106} (\bibinfo {year} {2001})}\BibitemShut {NoStop}%
\bibitem [{\citenamefont {Dyson}(1969)}]{Dyson:1968up}%
  \BibitemOpen
  \bibfield  {author} {\bibinfo {author} {\bibfnamefont {F.~J.}\ \bibnamefont
  {Dyson}},\ }\bibfield  {title} {\bibinfo {title} {{Existence of a phase
  transition in a one-dimensional Ising ferromagnet}},\ }\href
  {https://doi.org/10.1007/BF01645907} {\bibfield  {journal} {\bibinfo
  {journal} {Commun. Math. Phys.}\ }\textbf {\bibinfo {volume} {12}},\ \bibinfo
  {pages} {91} (\bibinfo {year} {1969})}\BibitemShut {NoStop}%
\bibitem [{\citenamefont {Defenu}\ \emph {et~al.}(2023)\citenamefont {Defenu},
  \citenamefont {Lerose},\ and\ \citenamefont
  {Pappalardi}}]{defenu2023outofequilibrium}%
  \BibitemOpen
  \bibfield  {author} {\bibinfo {author} {\bibfnamefont {N.}~\bibnamefont
  {Defenu}}, \bibinfo {author} {\bibfnamefont {A.}~\bibnamefont {Lerose}},\
  and\ \bibinfo {author} {\bibfnamefont {S.}~\bibnamefont {Pappalardi}},\
  }\href@noop {} {\bibinfo {title} {Out-of-equilibrium dynamics of quantum
  many-body systems with long-range interactions}} (\bibinfo {year} {2023}),\
  \Eprint {https://arxiv.org/abs/2307.04802} {arXiv:2307.04802
  [cond-mat.quant-gas]} \BibitemShut {NoStop}%
\bibitem [{\citenamefont {{Kitagawa}}\ and\ \citenamefont
  {{Ueda}}(1993)}]{Kitagawa1993}%
  \BibitemOpen
  \bibfield  {author} {\bibinfo {author} {\bibfnamefont {M.}~\bibnamefont
  {{Kitagawa}}}\ and\ \bibinfo {author} {\bibfnamefont {M.}~\bibnamefont
  {{Ueda}}},\ }\bibfield  {title} {\bibinfo {title} {{Squeezed spin states}},\
  }\href {https://doi.org/10.1103/PhysRevA.47.5138} {\bibfield  {journal}
  {\bibinfo  {journal} {\pra}\ }\textbf {\bibinfo {volume} {47}},\ \bibinfo
  {pages} {5138} (\bibinfo {year} {1993})}\BibitemShut {NoStop}%
\bibitem [{\citenamefont {Ballar~Trigueros}\ \emph {et~al.}(2023)\citenamefont
  {Ballar~Trigueros}, \citenamefont {Mendes~Santos},\ and\ \citenamefont
  {Heyl}}]{ballar_trigueros_fabian_2023_8252947}%
  \BibitemOpen
  \bibfield  {author} {\bibinfo {author} {\bibfnamefont {F.}~\bibnamefont
  {Ballar~Trigueros}}, \bibinfo {author} {\bibfnamefont {T.}~\bibnamefont
  {Mendes~Santos}},\ and\ \bibinfo {author} {\bibfnamefont {M.}~\bibnamefont
  {Heyl}},\ }\bibfield  {title} {\bibinfo {title} {{Mean-field theories are
  simple for neural quantum states}},\ }\href
  {https://doi.org/10.5281/zenodo.8252947} {10.5281/zenodo.8252947} (\bibinfo
  {year} {2023})\BibitemShut {NoStop}%
\bibitem [{\citenamefont {Weinberg}\ and\ \citenamefont
  {Bukov}(2017)}]{Weinberg_2017}%
  \BibitemOpen
  \bibfield  {author} {\bibinfo {author} {\bibfnamefont {P.}~\bibnamefont
  {Weinberg}}\ and\ \bibinfo {author} {\bibfnamefont {M.}~\bibnamefont
  {Bukov}},\ }\bibfield  {title} {\bibinfo {title} {{QuSpin}: a python package
  for dynamics and exact diagonalisation of quantum many body systems part i:
  spin chains},\ }\bibfield  {journal} {\bibinfo  {journal} {{SciPost}
  Physics}\ }\textbf {\bibinfo {volume} {2}},\ \href
  {https://doi.org/10.21468/scipostphys.2.1.003} {10.21468/scipostphys.2.1.003}
  (\bibinfo {year} {2017})\BibitemShut {NoStop}%
\bibitem [{\citenamefont {Schmitt}\ and\ \citenamefont
  {Reh}(2022)}]{Schmitt_2022}%
  \BibitemOpen
  \bibfield  {author} {\bibinfo {author} {\bibfnamefont {M.}~\bibnamefont
  {Schmitt}}\ and\ \bibinfo {author} {\bibfnamefont {M.}~\bibnamefont {Reh}},\
  }\bibfield  {title} {\bibinfo {title} {{jVMC}: Versatile and performant
  variational monte carlo leveraging automated differentiation and {GPU}
  acceleration},\ }\bibfield  {journal} {\bibinfo  {journal} {{SciPost} Physics
  Codebases}\ }\href {https://doi.org/10.21468/scipostphyscodeb.2}
  {10.21468/scipostphyscodeb.2} (\bibinfo {year} {2022})\BibitemShut {NoStop}%
\bibitem [{\citenamefont {Bradbury}\ \emph {et~al.}(2018)\citenamefont
  {Bradbury}, \citenamefont {Frostig}, \citenamefont {Hawkins}, \citenamefont
  {Johnson}, \citenamefont {Leary}, \citenamefont {Maclaurin}, \citenamefont
  {Necula}, \citenamefont {Paszke}, \citenamefont {Vander{P}las}, \citenamefont
  {Wanderman-{M}ilne},\ and\ \citenamefont {Zhang}}]{jax2018github}%
  \BibitemOpen
  \bibfield  {author} {\bibinfo {author} {\bibfnamefont {J.}~\bibnamefont
  {Bradbury}}, \bibinfo {author} {\bibfnamefont {R.}~\bibnamefont {Frostig}},
  \bibinfo {author} {\bibfnamefont {P.}~\bibnamefont {Hawkins}}, \bibinfo
  {author} {\bibfnamefont {M.~J.}\ \bibnamefont {Johnson}}, \bibinfo {author}
  {\bibfnamefont {C.}~\bibnamefont {Leary}}, \bibinfo {author} {\bibfnamefont
  {D.}~\bibnamefont {Maclaurin}}, \bibinfo {author} {\bibfnamefont
  {G.}~\bibnamefont {Necula}}, \bibinfo {author} {\bibfnamefont
  {A.}~\bibnamefont {Paszke}}, \bibinfo {author} {\bibfnamefont
  {J.}~\bibnamefont {Vander{P}las}}, \bibinfo {author} {\bibfnamefont
  {S.}~\bibnamefont {Wanderman-{M}ilne}},\ and\ \bibinfo {author}
  {\bibfnamefont {Q.}~\bibnamefont {Zhang}},\ }\href
  {http://github.com/google/jax} {\bibinfo {title} {{JAX}: composable
  transformations of {P}ython+{N}um{P}y programs}} (\bibinfo {year}
  {2018})\BibitemShut {NoStop}%
\bibitem [{\citenamefont {Popkov}\ \emph {et~al.}(2005)\citenamefont {Popkov},
  \citenamefont {Salerno},\ and\ \citenamefont {Schütz}}]{Popkov_2005}%
  \BibitemOpen
  \bibfield  {author} {\bibinfo {author} {\bibfnamefont {V.}~\bibnamefont
  {Popkov}}, \bibinfo {author} {\bibfnamefont {M.}~\bibnamefont {Salerno}},\
  and\ \bibinfo {author} {\bibfnamefont {G.}~\bibnamefont {Schütz}},\
  }\bibfield  {title} {\bibinfo {title} {Entangling power of
  permutation-invariant quantum states},\ }\bibfield  {journal} {\bibinfo
  {journal} {Physical Review A}\ }\textbf {\bibinfo {volume} {72}},\ \href
  {https://doi.org/10.1103/physreva.72.032327} {10.1103/physreva.72.032327}
  (\bibinfo {year} {2005})\BibitemShut {NoStop}%
\bibitem [{\citenamefont {Knap}\ \emph {et~al.}(2013)\citenamefont {Knap},
  \citenamefont {Abanin},\ and\ \citenamefont {Demler}}]{Knap_2013}%
  \BibitemOpen
  \bibfield  {author} {\bibinfo {author} {\bibfnamefont {M.}~\bibnamefont
  {Knap}}, \bibinfo {author} {\bibfnamefont {D.~A.}\ \bibnamefont {Abanin}},\
  and\ \bibinfo {author} {\bibfnamefont {E.}~\bibnamefont {Demler}},\
  }\bibfield  {title} {\bibinfo {title} {Dissipative dynamics of a driven
  quantum spin coupled to a bath of ultracold fermions},\ }\bibfield  {journal}
  {\bibinfo  {journal} {Physical Review Letters}\ }\textbf {\bibinfo {volume}
  {111}},\ \href {https://doi.org/10.1103/physrevlett.111.265302}
  {10.1103/physrevlett.111.265302} (\bibinfo {year} {2013})\BibitemShut
  {NoStop}%
\bibitem [{\citenamefont {Pan}\ and\ \citenamefont {Draayer}(1999)}]{PAN19991}%
  \BibitemOpen
  \bibfield  {author} {\bibinfo {author} {\bibfnamefont {F.}~\bibnamefont
  {Pan}}\ and\ \bibinfo {author} {\bibfnamefont {J.}~\bibnamefont {Draayer}},\
  }\bibfield  {title} {\bibinfo {title} {Analytical solutions for the lmg
  model},\ }\href
  {https://doi.org/https://doi.org/10.1016/S0370-2693(99)00191-4} {\bibfield
  {journal} {\bibinfo  {journal} {Physics Letters B}\ }\textbf {\bibinfo
  {volume} {451}},\ \bibinfo {pages} {1} (\bibinfo {year} {1999})}\BibitemShut
  {NoStop}%
\bibitem [{\citenamefont {Affonso}\ \emph {et~al.}(2022)\citenamefont
  {Affonso}, \citenamefont {Bissacot}, \citenamefont {Endo},\ and\
  \citenamefont {Handa}}]{affonso2022longrange}%
  \BibitemOpen
  \bibfield  {author} {\bibinfo {author} {\bibfnamefont {L.}~\bibnamefont
  {Affonso}}, \bibinfo {author} {\bibfnamefont {R.}~\bibnamefont {Bissacot}},
  \bibinfo {author} {\bibfnamefont {E.~O.}\ \bibnamefont {Endo}},\ and\
  \bibinfo {author} {\bibfnamefont {S.}~\bibnamefont {Handa}},\ }\href@noop {}
  {\bibinfo {title} {Long-range ising models: Contours, phase transitions and
  decaying fields}} (\bibinfo {year} {2022}),\ \Eprint
  {https://arxiv.org/abs/2105.06103} {arXiv:2105.06103 [math-ph]} \BibitemShut
  {NoStop}%
\bibitem [{\citenamefont {Kac}\ and\ \citenamefont
  {Thompson}(1969)}]{doi:10.1063/1.1664976}%
  \BibitemOpen
  \bibfield  {author} {\bibinfo {author} {\bibfnamefont {M.}~\bibnamefont
  {Kac}}\ and\ \bibinfo {author} {\bibfnamefont {C.~J.}\ \bibnamefont
  {Thompson}},\ }\bibfield  {title} {\bibinfo {title} {Critical behavior of
  several lattice models with long range interaction},\ }\href
  {https://doi.org/10.1063/1.1664976} {\bibfield  {journal} {\bibinfo
  {journal} {Journal of Mathematical Physics}\ }\textbf {\bibinfo {volume}
  {10}},\ \bibinfo {pages} {1373} (\bibinfo {year} {1969})},\ \Eprint
  {https://arxiv.org/abs/https://doi.org/10.1063/1.1664976}
  {https://doi.org/10.1063/1.1664976} \BibitemShut {NoStop}%
\bibitem [{\citenamefont {Nomura}(2022)}]{doi:10.7566/JPSJ.91.054709}%
  \BibitemOpen
  \bibfield  {author} {\bibinfo {author} {\bibfnamefont {Y.}~\bibnamefont
  {Nomura}},\ }\bibfield  {title} {\bibinfo {title} {Investigating network
  parameters in neural-network quantum states},\ }\href
  {https://doi.org/10.7566/JPSJ.91.054709} {\bibfield  {journal} {\bibinfo
  {journal} {Journal of the Physical Society of Japan}\ }\textbf {\bibinfo
  {volume} {91}},\ \bibinfo {pages} {054709} (\bibinfo {year} {2022})},\
  \Eprint {https://arxiv.org/abs/https://doi.org/10.7566/JPSJ.91.054709}
  {https://doi.org/10.7566/JPSJ.91.054709} \BibitemShut {NoStop}%
\bibitem [{\citenamefont {Levine}\ \emph {et~al.}(2019)\citenamefont {Levine},
  \citenamefont {Sharir}, \citenamefont {Cohen},\ and\ \citenamefont
  {Shashua}}]{PhysRevLett.122.065301}%
  \BibitemOpen
  \bibfield  {author} {\bibinfo {author} {\bibfnamefont {Y.}~\bibnamefont
  {Levine}}, \bibinfo {author} {\bibfnamefont {O.}~\bibnamefont {Sharir}},
  \bibinfo {author} {\bibfnamefont {N.}~\bibnamefont {Cohen}},\ and\ \bibinfo
  {author} {\bibfnamefont {A.}~\bibnamefont {Shashua}},\ }\bibfield  {title}
  {\bibinfo {title} {Quantum entanglement in deep learning architectures},\
  }\href {https://doi.org/10.1103/PhysRevLett.122.065301} {\bibfield  {journal}
  {\bibinfo  {journal} {Phys. Rev. Lett.}\ }\textbf {\bibinfo {volume} {122}},\
  \bibinfo {pages} {065301} (\bibinfo {year} {2019})}\BibitemShut {NoStop}%
\bibitem [{\citenamefont {Jin}\ \emph {et~al.}(2020)\citenamefont {Jin},
  \citenamefont {Yi}, \citenamefont {Zhang}, \citenamefont {Zhang},
  \citenamefont {Schewe},\ and\ \citenamefont {Huang}}]{jin2020does}%
  \BibitemOpen
  \bibfield  {author} {\bibinfo {author} {\bibfnamefont {G.}~\bibnamefont
  {Jin}}, \bibinfo {author} {\bibfnamefont {X.}~\bibnamefont {Yi}}, \bibinfo
  {author} {\bibfnamefont {L.}~\bibnamefont {Zhang}}, \bibinfo {author}
  {\bibfnamefont {L.}~\bibnamefont {Zhang}}, \bibinfo {author} {\bibfnamefont
  {S.}~\bibnamefont {Schewe}},\ and\ \bibinfo {author} {\bibfnamefont
  {X.}~\bibnamefont {Huang}},\ }\href@noop {} {\bibinfo {title} {How does
  weight correlation affect the generalisation ability of deep neural
  networks}} (\bibinfo {year} {2020}),\ \Eprint
  {https://arxiv.org/abs/2010.05983} {arXiv:2010.05983 [cs.LG]} \BibitemShut
  {NoStop}%
\bibitem [{\citenamefont {Wu}\ \emph {et~al.}(2023)\citenamefont {Wu},
  \citenamefont {Rossi}, \citenamefont {Vicentini}, \citenamefont
  {Astrakhantsev}, \citenamefont {Becca}, \citenamefont {Cao}, \citenamefont
  {Carrasquilla}, \citenamefont {Ferrari}, \citenamefont {Georges},
  \citenamefont {Hibat-Allah}, \citenamefont {Imada}, \citenamefont {Läuchli},
  \citenamefont {Mazzola}, \citenamefont {Mezzacapo}, \citenamefont {Millis},
  \citenamefont {Moreno}, \citenamefont {Neupert}, \citenamefont {Nomura},
  \citenamefont {Nys}, \citenamefont {Parcollet}, \citenamefont {Pohle},
  \citenamefont {Romero}, \citenamefont {Schmid}, \citenamefont {Silvester},
  \citenamefont {Sorella}, \citenamefont {Tocchio}, \citenamefont {Wang},
  \citenamefont {White}, \citenamefont {Wietek}, \citenamefont {Yang},
  \citenamefont {Yang}, \citenamefont {Zhang},\ and\ \citenamefont
  {Carleo}}]{wu2023variational}%
  \BibitemOpen
  \bibfield  {author} {\bibinfo {author} {\bibfnamefont {D.}~\bibnamefont
  {Wu}}, \bibinfo {author} {\bibfnamefont {R.}~\bibnamefont {Rossi}}, \bibinfo
  {author} {\bibfnamefont {F.}~\bibnamefont {Vicentini}}, \bibinfo {author}
  {\bibfnamefont {N.}~\bibnamefont {Astrakhantsev}}, \bibinfo {author}
  {\bibfnamefont {F.}~\bibnamefont {Becca}}, \bibinfo {author} {\bibfnamefont
  {X.}~\bibnamefont {Cao}}, \bibinfo {author} {\bibfnamefont {J.}~\bibnamefont
  {Carrasquilla}}, \bibinfo {author} {\bibfnamefont {F.}~\bibnamefont
  {Ferrari}}, \bibinfo {author} {\bibfnamefont {A.}~\bibnamefont {Georges}},
  \bibinfo {author} {\bibfnamefont {M.}~\bibnamefont {Hibat-Allah}}, \bibinfo
  {author} {\bibfnamefont {M.}~\bibnamefont {Imada}}, \bibinfo {author}
  {\bibfnamefont {A.~M.}\ \bibnamefont {Läuchli}}, \bibinfo {author}
  {\bibfnamefont {G.}~\bibnamefont {Mazzola}}, \bibinfo {author} {\bibfnamefont
  {A.}~\bibnamefont {Mezzacapo}}, \bibinfo {author} {\bibfnamefont
  {A.}~\bibnamefont {Millis}}, \bibinfo {author} {\bibfnamefont {J.~R.}\
  \bibnamefont {Moreno}}, \bibinfo {author} {\bibfnamefont {T.}~\bibnamefont
  {Neupert}}, \bibinfo {author} {\bibfnamefont {Y.}~\bibnamefont {Nomura}},
  \bibinfo {author} {\bibfnamefont {J.}~\bibnamefont {Nys}}, \bibinfo {author}
  {\bibfnamefont {O.}~\bibnamefont {Parcollet}}, \bibinfo {author}
  {\bibfnamefont {R.}~\bibnamefont {Pohle}}, \bibinfo {author} {\bibfnamefont
  {I.}~\bibnamefont {Romero}}, \bibinfo {author} {\bibfnamefont
  {M.}~\bibnamefont {Schmid}}, \bibinfo {author} {\bibfnamefont {J.~M.}\
  \bibnamefont {Silvester}}, \bibinfo {author} {\bibfnamefont {S.}~\bibnamefont
  {Sorella}}, \bibinfo {author} {\bibfnamefont {L.~F.}\ \bibnamefont
  {Tocchio}}, \bibinfo {author} {\bibfnamefont {L.}~\bibnamefont {Wang}},
  \bibinfo {author} {\bibfnamefont {S.~R.}\ \bibnamefont {White}}, \bibinfo
  {author} {\bibfnamefont {A.}~\bibnamefont {Wietek}}, \bibinfo {author}
  {\bibfnamefont {Q.}~\bibnamefont {Yang}}, \bibinfo {author} {\bibfnamefont
  {Y.}~\bibnamefont {Yang}}, \bibinfo {author} {\bibfnamefont {S.}~\bibnamefont
  {Zhang}},\ and\ \bibinfo {author} {\bibfnamefont {G.}~\bibnamefont
  {Carleo}},\ }\href@noop {} {\bibinfo {title} {Variational benchmarks for
  quantum many-body problems}} (\bibinfo {year} {2023}),\ \Eprint
  {https://arxiv.org/abs/2302.04919} {arXiv:2302.04919 [quant-ph]} \BibitemShut
  {NoStop}%
\bibitem [{\citenamefont {Granet}(2023)}]{10.21468/SciPostPhys.14.5.133}%
  \BibitemOpen
  \bibfield  {author} {\bibinfo {author} {\bibfnamefont {E.}~\bibnamefont
  {Granet}},\ }\bibfield  {title} {\bibinfo {title} {{Exact mean-field solution
  of a spin chain with short-range and long-range interactions}},\ }\href
  {https://doi.org/10.21468/SciPostPhys.14.5.133} {\bibfield  {journal}
  {\bibinfo  {journal} {SciPost Phys.}\ }\textbf {\bibinfo {volume} {14}},\
  \bibinfo {pages} {133} (\bibinfo {year} {2023})}\BibitemShut {NoStop}%
\bibitem [{\citenamefont {Casagrande}\ \emph {et~al.}(2023)\citenamefont
  {Casagrande}, \citenamefont {Xing}, \citenamefont {Dalmonte}, \citenamefont
  {Rodriguez}, \citenamefont {Balachandran},\ and\ \citenamefont
  {Poletti}}]{casagrande2023complexity}%
  \BibitemOpen
  \bibfield  {author} {\bibinfo {author} {\bibfnamefont {H.~P.}\ \bibnamefont
  {Casagrande}}, \bibinfo {author} {\bibfnamefont {B.}~\bibnamefont {Xing}},
  \bibinfo {author} {\bibfnamefont {M.}~\bibnamefont {Dalmonte}}, \bibinfo
  {author} {\bibfnamefont {A.}~\bibnamefont {Rodriguez}}, \bibinfo {author}
  {\bibfnamefont {V.}~\bibnamefont {Balachandran}},\ and\ \bibinfo {author}
  {\bibfnamefont {D.}~\bibnamefont {Poletti}},\ }\href@noop {} {\bibinfo
  {title} {Complexity of spin configurations dynamics due to unitary evolution
  and periodic projective measurements}} (\bibinfo {year} {2023}),\ \Eprint
  {https://arxiv.org/abs/2305.03334} {arXiv:2305.03334 [cond-mat.stat-mech]}
  \BibitemShut {NoStop}%
\bibitem [{\citenamefont {Fey}\ and\ \citenamefont
  {Schmidt}(2016)}]{PhysRevB.94.075156}%
  \BibitemOpen
  \bibfield  {author} {\bibinfo {author} {\bibfnamefont {S.}~\bibnamefont
  {Fey}}\ and\ \bibinfo {author} {\bibfnamefont {K.~P.}\ \bibnamefont
  {Schmidt}},\ }\bibfield  {title} {\bibinfo {title} {Critical behavior of
  quantum magnets with long-range interactions in the thermodynamic limit},\
  }\href {https://doi.org/10.1103/PhysRevB.94.075156} {\bibfield  {journal}
  {\bibinfo  {journal} {Phys. Rev. B}\ }\textbf {\bibinfo {volume} {94}},\
  \bibinfo {pages} {075156} (\bibinfo {year} {2016})}\BibitemShut {NoStop}%
\bibitem [{\citenamefont {Fisher}\ \emph {et~al.}(1972)\citenamefont {Fisher},
  \citenamefont {Ma},\ and\ \citenamefont {Nickel}}]{PhysRevLett.29.917}%
  \BibitemOpen
  \bibfield  {author} {\bibinfo {author} {\bibfnamefont {M.~E.}\ \bibnamefont
  {Fisher}}, \bibinfo {author} {\bibfnamefont {S.-k.}\ \bibnamefont {Ma}},\
  and\ \bibinfo {author} {\bibfnamefont {B.~G.}\ \bibnamefont {Nickel}},\
  }\bibfield  {title} {\bibinfo {title} {Critical exponents for long-range
  interactions},\ }\href {https://doi.org/10.1103/PhysRevLett.29.917}
  {\bibfield  {journal} {\bibinfo  {journal} {Phys. Rev. Lett.}\ }\textbf
  {\bibinfo {volume} {29}},\ \bibinfo {pages} {917} (\bibinfo {year}
  {1972})}\BibitemShut {NoStop}%
\bibitem [{\citenamefont {Vodola}\ \emph {et~al.}(2015)\citenamefont {Vodola},
  \citenamefont {Lepori}, \citenamefont {Ercolessi},\ and\ \citenamefont
  {Pupillo}}]{Vodola_2015}%
  \BibitemOpen
  \bibfield  {author} {\bibinfo {author} {\bibfnamefont {D.}~\bibnamefont
  {Vodola}}, \bibinfo {author} {\bibfnamefont {L.}~\bibnamefont {Lepori}},
  \bibinfo {author} {\bibfnamefont {E.}~\bibnamefont {Ercolessi}},\ and\
  \bibinfo {author} {\bibfnamefont {G.}~\bibnamefont {Pupillo}},\ }\bibfield
  {title} {\bibinfo {title} {Long-range ising and kitaev models: phases,
  correlations and edge modes},\ }\href
  {https://doi.org/10.1088/1367-2630/18/1/015001} {\bibfield  {journal}
  {\bibinfo  {journal} {New Journal of Physics}\ }\textbf {\bibinfo {volume}
  {18}},\ \bibinfo {pages} {015001} (\bibinfo {year} {2015})}\BibitemShut
  {NoStop}%
\bibitem [{\citenamefont {Zhang}\ \emph {et~al.}(2017)\citenamefont {Zhang},
  \citenamefont {Pagano}, \citenamefont {Hess}, \citenamefont {Kyprianidis},
  \citenamefont {Becker}, \citenamefont {Kaplan}, \citenamefont {Gorshkov},
  \citenamefont {Gong},\ and\ \citenamefont {Monroe}}]{Zhang_2017}%
  \BibitemOpen
  \bibfield  {author} {\bibinfo {author} {\bibfnamefont {J.}~\bibnamefont
  {Zhang}}, \bibinfo {author} {\bibfnamefont {G.}~\bibnamefont {Pagano}},
  \bibinfo {author} {\bibfnamefont {P.~W.}\ \bibnamefont {Hess}}, \bibinfo
  {author} {\bibfnamefont {A.}~\bibnamefont {Kyprianidis}}, \bibinfo {author}
  {\bibfnamefont {P.}~\bibnamefont {Becker}}, \bibinfo {author} {\bibfnamefont
  {H.}~\bibnamefont {Kaplan}}, \bibinfo {author} {\bibfnamefont {A.~V.}\
  \bibnamefont {Gorshkov}}, \bibinfo {author} {\bibfnamefont {Z.-X.}\
  \bibnamefont {Gong}},\ and\ \bibinfo {author} {\bibfnamefont
  {C.}~\bibnamefont {Monroe}},\ }\bibfield  {title} {\bibinfo {title}
  {Observation of a many-body dynamical phase transition with a 53-qubit
  quantum simulator},\ }\href {https://doi.org/10.1038/nature24654} {\bibfield
  {journal} {\bibinfo  {journal} {Nature}\ }\textbf {\bibinfo {volume} {551}},\
  \bibinfo {pages} {601} (\bibinfo {year} {2017})}\BibitemShut {NoStop}%
\bibitem [{\citenamefont {Tran}\ \emph {et~al.}(2021)\citenamefont {Tran},
  \citenamefont {Guo}, \citenamefont {Baldwin}, \citenamefont {Ehrenberg},
  \citenamefont {Gorshkov},\ and\ \citenamefont {Lucas}}]{Tran_2021}%
  \BibitemOpen
  \bibfield  {author} {\bibinfo {author} {\bibfnamefont {M.~C.}\ \bibnamefont
  {Tran}}, \bibinfo {author} {\bibfnamefont {A.~Y.}\ \bibnamefont {Guo}},
  \bibinfo {author} {\bibfnamefont {C.~L.}\ \bibnamefont {Baldwin}}, \bibinfo
  {author} {\bibfnamefont {A.}~\bibnamefont {Ehrenberg}}, \bibinfo {author}
  {\bibfnamefont {A.~V.}\ \bibnamefont {Gorshkov}},\ and\ \bibinfo {author}
  {\bibfnamefont {A.}~\bibnamefont {Lucas}},\ }\bibfield  {title} {\bibinfo
  {title} {Lieb-robinson light cone for power-law interactions},\ }\bibfield
  {journal} {\bibinfo  {journal} {Physical Review Letters}\ }\textbf {\bibinfo
  {volume} {127}},\ \href {https://doi.org/10.1103/physrevlett.127.160401}
  {10.1103/physrevlett.127.160401} (\bibinfo {year} {2021})\BibitemShut
  {NoStop}%
\bibitem [{\citenamefont {Eldredge}\ \emph {et~al.}(2017)\citenamefont
  {Eldredge}, \citenamefont {Gong}, \citenamefont {Young}, \citenamefont
  {Moosavian}, \citenamefont {Foss-Feig},\ and\ \citenamefont
  {Gorshkov}}]{Eldredge_2017}%
  \BibitemOpen
  \bibfield  {author} {\bibinfo {author} {\bibfnamefont {Z.}~\bibnamefont
  {Eldredge}}, \bibinfo {author} {\bibfnamefont {Z.-X.}\ \bibnamefont {Gong}},
  \bibinfo {author} {\bibfnamefont {J.~T.}\ \bibnamefont {Young}}, \bibinfo
  {author} {\bibfnamefont {A.~H.}\ \bibnamefont {Moosavian}}, \bibinfo {author}
  {\bibfnamefont {M.}~\bibnamefont {Foss-Feig}},\ and\ \bibinfo {author}
  {\bibfnamefont {A.~V.}\ \bibnamefont {Gorshkov}},\ }\bibfield  {title}
  {\bibinfo {title} {Fast quantum state transfer and entanglement
  renormalization using long-range interactions},\ }\bibfield  {journal}
  {\bibinfo  {journal} {Physical Review Letters}\ }\textbf {\bibinfo {volume}
  {119}},\ \href {https://doi.org/10.1103/physrevlett.119.170503}
  {10.1103/physrevlett.119.170503} (\bibinfo {year} {2017})\BibitemShut
  {NoStop}%
\bibitem [{\citenamefont {Park}\ and\ \citenamefont
  {Kastoryano}(2020)}]{Park_2020}%
  \BibitemOpen
  \bibfield  {author} {\bibinfo {author} {\bibfnamefont {C.-Y.}\ \bibnamefont
  {Park}}\ and\ \bibinfo {author} {\bibfnamefont {M.~J.}\ \bibnamefont
  {Kastoryano}},\ }\bibfield  {title} {\bibinfo {title} {Geometry of learning
  neural quantum states},\ }\bibfield  {journal} {\bibinfo  {journal} {Physical
  Review Research}\ }\textbf {\bibinfo {volume} {2}},\ \href
  {https://doi.org/10.1103/physrevresearch.2.023232}
  {10.1103/physrevresearch.2.023232} (\bibinfo {year} {2020})\BibitemShut
  {NoStop}%
\bibitem [{\citenamefont {Russomanno}\ \emph {et~al.}(2021)\citenamefont
  {Russomanno}, \citenamefont {Fava},\ and\ \citenamefont
  {Heyl}}]{PhysRevB.104.094309}%
  \BibitemOpen
  \bibfield  {author} {\bibinfo {author} {\bibfnamefont {A.}~\bibnamefont
  {Russomanno}}, \bibinfo {author} {\bibfnamefont {M.}~\bibnamefont {Fava}},\
  and\ \bibinfo {author} {\bibfnamefont {M.}~\bibnamefont {Heyl}},\ }\bibfield
  {title} {\bibinfo {title} {Quantum chaos and ensemble inequivalence of
  quantum long-range ising chains},\ }\href
  {https://doi.org/10.1103/PhysRevB.104.094309} {\bibfield  {journal} {\bibinfo
   {journal} {Phys. Rev. B}\ }\textbf {\bibinfo {volume} {104}},\ \bibinfo
  {pages} {094309} (\bibinfo {year} {2021})}\BibitemShut {NoStop}%
\end{thebibliography}%

\onecolumngrid

\appendix

\section{\label{Ap: Analytic} Analytical computations} 

\subsection{Ground state energy}

In this appendix, we provide further details on the analytical computations discussed throughout the text. We start by recalling our model Hamiltonian and normalization factor:

\begin{equation}
    H = \frac{-J}{\mathcal{N}(L,\alpha)} \sum_{i\neq j}^{} \frac{1}{|i-j|^{\alpha}} S^{z}_{i} S^{z}_{j}  - g \sum_{i}^{} S^{x}_{i},
\end{equation}

\begin{equation}
    \mathcal{N}(L,\alpha) = \frac{1}{L-1} \sum_{i\neq j} \frac{1}{|i-j|^{\alpha}}.
\end{equation}

We are interested in computing the local energy, which is defined as $E_{\textrm{loc}}(s) = \frac{\langle s | H | \Psi \rangle}{\langle s | \Psi \rangle}$. For our model, the local energy takes the form:

\begin{equation}
    E_{\textrm{loc}}(s) = \frac{-J}{\mathcal{N}(L,\alpha)} \sum_{i\neq j}^{} \frac{s_{i} s_{j}}{|i-j|^{\alpha}}   - g \sum_{i}^{} \frac{\langle s'(s_i) | \Psi \rangle}{\langle s | \Psi \rangle},
\end{equation}

\noindent
where $s'(s_i)$ means we have flipped the spin in the $i$-th site. Using the fact that $\log \langle s | \Psi \rangle = \log \cosh \left( W \sum_{i}^{} s_i\right)$ from our ansatz, we can rewrite the second term as:

\begin{equation}
    E_{\textrm{loc}}(s) = \frac{-J}{\mathcal{N}(L,\alpha)} \sum_{i\neq j}^{} \frac{s_{i} s_{j}}{|i-j|^{\alpha}}   - g \sum_{i}^{} \exp \left\{ \log \cosh \left( W \sum_{i}^{} s'_i\right) - \log \cosh \left( W \sum_{i}^{} s_i\right)  \right\}.
\end{equation}

In the ferromagnetic phase, we can use the approximation $\log \cosh x \approx x$ for $x \sim L \gg 1$. Making this approximation, the second term reduces to:

\begin{equation}
    E_{\textrm{loc}}(s) = \frac{-J}{\mathcal{N}(L,\alpha)} \sum_{i\neq j}^{} \frac{s_{i} s_{j}}{|i-j|^{\alpha}}   - g \sum_{i}^{} e^{-2Ws_i}.
\end{equation}

To eliminate the exponential term and have all dependence on $s_i$ be polynomial, we can make use of the following identity:

\begin{equation}
    e^{-2W s_i} = \cosh(2W) - \sinh(2W) s_i, \  \textrm{for } s_i = \pm 1.
\end{equation}

This simplification may not be immediately obvious, but it will be helpful when we sum over all spin configurations. Using this identity, we can simplify our local energy expression to the form:

\begin{equation}
    E_{\textrm{loc}}(s) = \frac{-J}{\mathcal{N}(L,\alpha)} \sum_{i\neq j}^{} \frac{s_{i} s_{j}}{|i-j|^{\alpha}}   - g L \cosh(2W) + g \sinh(2W) \sum_{i}^{} s_i.
\end{equation}

With this local energy, we can compute the total energy of the system as follows:

\begin{equation}
    E = \sum_{\mathbf{S}}^{} E_{\textrm{loc}}(s) \frac{|\langle s | \Psi \rangle|^2}{N}; \ \ \  N \equiv \sum_{\mathbf{S}}^{} |\langle s | \Psi \rangle|^2.
\end{equation}

The normalization can be readily computed from our ansatz and the exponential term simplification,

\begin{equation}
    N = |\langle s | \Psi \rangle|^2 = \sum_{\mathbf{S}}^{} e^{2W\sum_{i}^{} s_i}  = \sum_{\mathbf{S}}^{} \prod_i e^{2W s_i} =\sum_{\mathbf{S}}^{} \prod_i \cosh(2W) - g \sinh(2W) s_i = \cosh^{L}(2W) 2^{L}.
\end{equation}

Analogously, by substituting into the equation for the total energy, we derive the subsequent expression:

\begin{multline}
        E = \frac{1}{\cosh^{L}(2W) 2^{L}} \Bigg[ \frac{-J}{\mathcal{N}(L,\alpha)} \sum_{i\neq j} 
 \frac{1}{|i-j|^{\alpha}} \sum_{\mathbf{S}}^{} s_i  s_j \prod_{k} \left( \cosh(2W) +  \sinh(2W) s_k \right) \\ + g \sinh(2W) \sum_{\mathbf{S}}^{} \sum_{i}^{} s_i \prod_{k} \left( \cosh(2W) +  \sinh(2W) s_k \right) \Bigg]  - gL\cosh(2W)
\end{multline}

The importance of the polynomial dependence in $s_i$ becomes evident in this expression. Specifically, in the first term, two terms in the product will interact with the external $s_i$ and $s_j$, respectively, namely $\left( \cosh(2W)s_i + \sinh(2W) \right)$ and $\left( \cosh(2W)s_j + \sinh(2W) \right)$. Thus, the first term contains two terms that give $\sinh(2W)$ and $L-2$ terms that give $\cosh(2W)$, so that the overall expression simplifies to:

\begin{multline}
E = \frac{1}{\cosh^{L}(2W) 2^{L}} \Bigg[ \frac{-J}{\mathcal{N}(L,\alpha)} \sum_{i\neq j}
\frac{1}{|i-j|^{\alpha}} 2^L \sinh^2(2W) \cosh^{L-2}(2W) \ + g \sinh(2W) \sum_{i}^{} 2^L \sinh(2W) \cosh^{L-1}(2W) \Bigg]\\ - gL\cosh(2W)
\end{multline}

Having successfully summed over all spin configurations, we arrive at the final expression for the energy:

\begin{equation}
E = -J(L-1) \tanh^{2}(2W) + gL\tanh(2W) \sinh(2W) -gL\cosh(2W).
\label{eq:lHr}
\end{equation}

The minimization procedure is straightforward from here:

\begin{equation}
\frac{\partial E}{\partial W} = 0 = \left[ 1 - \tanh^{2}(2W) \right] \ \underbrace{ \left[ -2J(L-1) \tanh(2W) + gL\sinh(2W) \right]}_{\cosh(2W) = \frac{2J}{gL}(L-1)}.
\end{equation}

\subsection{Energy fluctuations}

The other quantity we need to compute analytically is the energy fluctuation density,

\begin{equation}
    \sigma (H)^2 = \frac{1}{L} \Big[\langle H^2 \rangle - \langle H \rangle^2 \Big].
\end{equation}

For the fluctuations, all that remains to be computed is the first term $\langle H^2 \rangle$ since the second term is precisely the energy computed above $ \langle H \rangle = E$. In particular, a good simplification we can perform to rewrite the $H^2$ term in terms of the local energy is:

\begin{equation}
    \langle H^2 \rangle = \frac{1}{N} \sum_{\mathbf{S}}^{} E_{\textrm{loc}}(s)^2 \ |\langle s | \Psi \rangle|^2.
\end{equation}

The form of $E_{\textrm{loc}}(s)^2$ is given by:

\begin{multline}
    E_{\textrm{loc}}(s)^2 = \frac{J^2}{\mathcal{N}(L,\alpha)^2} \sum_{i \neq j, n \neq k} \frac{s_i s_j s_n s_k}{|i-j|^{\alpha} |n-k|^{\alpha}} + \frac{2 J g L}{\mathcal{N}(L,\alpha)} \cosh(2W) \sum_{i \neq j} \frac{s_i s_j}{|i-j|^{\alpha}} + g^2 L^2 \cosh^2(2W) \\ - \frac{2 J g}{\mathcal{N}(L,\alpha)} \sinh(2W) \sum_{i \neq j, k} \frac{s_i s_j s_k}{|i-j|^{\alpha}} - 2 g^2 L \cosh(2W) \sinh(2W) \sum_{i} s_i + g^2 \sinh^2(2W) \sum_{i \neq j} s_i s_j.
    \label{eq:Elo2}
\end{multline}

Now, to sum over spin configurations it is convenient to split the sums into sums where all indices are different. To give an example of this consider the expansion of the first term in Eq.\,(\ref{eq:Elo2}), 

\begin{multline}
    \sum_{i \neq j, n \neq k} \frac{s_i s_j s_n s_k}{|i-j|^{\alpha} |n-k|^{\alpha}} = \sum_{i \neq j \neq n \neq k} \frac{s_i s_j s_n s_k}{|i-j|^{\alpha} |n-k|^{\alpha}} + 4 \sum_{i \neq j \neq k} \frac{s_i s_k}{|i-j|^{\alpha} |j-k|^{\alpha}} + 2 \sum_{i \neq j} \frac{1}{|i-j|^{2 \alpha}}.
    \label{eq:Ident}
\end{multline}

Analogously, we may rewrite all sums in this way to apply the same method to sum over spin configurations as the one used in the energy calculation. If one follows this process we find that the value for the expectation value of $H^2$ is:

\begin{multline}
    \langle H^2 \rangle  = \frac{J^2}{\mathcal{N}(L,\alpha)^2} \left[ \tanh^4(2W)  \sum_{i \neq j \neq n \neq k} \frac{1}{|i-j|^{\alpha} |n-k|^{\alpha}}  + 4 \tanh^2(2W) \sum_{i \neq j \neq k} \frac{1}{|i-j|^{\alpha} |j-k|^{\alpha}} + 2 \sum_{i \neq j} \frac{1}{|i-j|^{2 \alpha}}  \right] \\ + 2 J g L (L-1) \cosh(2W) \tanh^2(2W) - 2 J g (L-1) \sinh(2W) \left[ \tanh^3(2W) (L-2)  + 2 \tanh(2W) \right] + g^2 L^2 \cosh^2(2W) \\ -2g^2L^2\sinh^2(2W) + g^2 \sinh^2(2W) \tanh^2(2W) L^2 - g^2  \sinh^2(2W) \tanh^2(2W) L + L g^2 \sinh^2(2W).
    \label{eq:lH2r}
\end{multline}

With equations (\ref{eq:lHr}) and (\ref{eq:lH2r}) one can now compute the fluctuations directly as a function of the variational parameter $W$. To reach the specific expression in Eq.\,(\ref{eq:FluctAlp}) we reuse the identity in (\ref{eq:Ident}) but now to get rid of the $4$ index sum. It is also convenient to write all the sums in terms of the generalized harmonic numbers. To do that we use the following two identities for $L$ odd\,(Even case can be obtained by subtracting a correction term due to over-counting):

\begin{equation}
    \sum_{i\neq j} \frac{1}{|i-j|^{\alpha}} = 
    \displaystyle 2L \, H_{\lfloor\frac{L}{2}\rfloor,\alpha},
\end{equation}

\begin{equation}
    \sum_{i\neq j\neq k} \frac{1}{|i-j|^{\alpha}|j-k|^{\alpha}} = L \, \left( 4 H_{\lfloor\frac{L}{2}\rfloor,\alpha}^2 - 2 H_{\lfloor\frac{L}{2}\rfloor,2\alpha}\right).
\end{equation}

\subsection{Magnetization}
To compute the magnetization as expressed in Eq.\,(\ref{eq:Magnetization}), we calculate the expected value of the $S^{z}_{i}$ operator with the one parameter ansatz. The calculation is performed as follows:
\begin{equation}
    \langle  S^{z}_i  \rangle = \frac{1}{N} \sum_{\mathbf{S}}^{} s_i \ |\langle s | \Psi \rangle|^2 = \frac{1}{N} \sum_{\mathbf{S}}^{} s_i \ \prod_{k} \left( \cosh(2W) +  \sinh(2W) s_k \right)= \tanh(2W).
\end{equation}

\end{document}